\documentclass[11pt, oneside]{article}   	% use "amsart" instead of "article" for AMSLaTeX format
\usepackage{geometry}
\usepackage{xcolor} 
\usepackage{amsmath}
\usepackage{amsfonts}
\usepackage{setspace}
\usepackage{amssymb}
\usepackage{mathrsfs}
\usepackage{bbold}
\usepackage{cancel}
\usepackage[T1]{fontenc}
\usepackage{slashed}
\usepackage[most]{tcolorbox}
\usepackage{array, bigdelim,multirow,multicol}
\usepackage[toc,page]{appendix}
\usepackage{dsfont}
\usepackage{longtable}
\usepackage{ltablex}
\usepackage{enumitem}
\usepackage{cite}
\usepackage{jheppub}
\usepackage[framemethod=tikz]{mdframed}
\usepackage{environ}
\NewEnviron{myalign}{%
\begin{align}
\scalebox{0.9}{$\BODY$}
\end{align}
}
\usepackage{graphicx}				% Use pdf, png, jpg, or eps§ with pdflatex; use eps in DVI mode							% TeX will automatically convert eps --> pdf in pdflate
\usepackage{bbm}
\usepackage{bm}	
\usepackage[normalem]{ulem}							
\newsavebox\myv

\newcommand{\ignore}[1]{}
\newcommand{\dd}{\displaystyle}

\newcommand{\eps}{\epsilon}

\newcommand{\nn}{\nonumber}
\newcommand{\be}{\begin{equation}}
\newcommand{\ee}{\end{equation}}
\newcommand{\bea}{\begin{eqnarray}}
\newcommand{\eea}{\end{eqnarray}}

\def\gapmatrix[{\begin{pmatrix}
\gaprows}
\def\gaprows#1[#2]#3{%
\gapcell#2\gapendrow,\ifx]#3\end{pmatrix}\else\afterfi\\\gaprows\fi}

\def\afterfi#1\fi{\fi#1}
\def\gapcell#1,{#1\uppercase{&}\gapcell}
\def\gapendrow#1\gapcell{}

\allowdisplaybreaks

\makeatletter
\renewcommand*{\@fnsymbol}[1]{\ensuremath{\ifcase#1\or *\or  \mathsection\or \ddagger\or
\dagger\or \mathparagraph\or \|\or **\or \dagger\dagger
\or \ddagger\ddagger \else\@ctrerr\fi}}
\makeatother

\newcommand{\MeV}{{\rm \, MeV}}

\numberwithin{equation}{section}

\title{\bf 
On the decay of a light spinless particle\\
into two photons}

\abstract{We analyze the effective couplings of a light, spinless, gauge-singlet particle $\phi$ to on-shell photons.
Starting from the most general theory at the electroweak scale, which allows for CP-violating interactions
suppressed by inverse powers of an ultraviolet scale $\Lambda$, we derive the corresponding low-energy effective theory valid below the GeV scale.
Within this framework, we systematically expand the effective couplings of $\phi$ to on-shell photons
in powers of small parameters. Working at the one-loop level, we retain terms at first order in $1/\Lambda$.
We incorporate both isospin-breaking effects and $\eta$--$\eta'$ mixing, and provide explicit expressions
for the couplings up to first order in $m_\phi^2/m_\eta^2$ and $m_\pi^2/m_\eta^2$.
As applications, we compute the decay rate of $\phi$ into two photons and illustrate our results in several
physically motivated scenarios.}

\preprint{TTP25-057, P3H-25-118}

\author[a]{Ferruccio Feruglio,}
\author[b]{Gabriele Levati,}
\author[c]{Robert Ziegler}

\affiliation{
$^a$ INFN, Sezione di Padova, Via Marzolo 8, I-35131 Padua, Italy \\
$^b$ Albert Einstein Center for Fundamental Physics, Institute for Theoretical Physics, \\
University of Bern, Sidlerstrasse 5, 3012 Bern, Switzerland\\ 
$^c$ Institute for Theoretical Particle Physics (TTP), Karlsruhe Institute of Technology (KIT), \\ Wolfgang-Gaede-Str. 1, D-76131 Karlsruhe, Germany\\
}

\begin{document}
\maketitle
\noindent
\section{Introduction}

The absence of any unambiguous evidence for heavy new physics (NP) at the LHC has motivated growing interest -- both theoretical and experimental -- towards alternative scenarios featuring new light particles with masses much lighter than a proton.  As the low-energy remnants of a possibly rich dynamics above the electroweak scale, light NP candidates would offer the unique opportunity to probe new interactions beyond the SM ones. 
A prominent example in this direction is given by light spinless particles, whose mass can be naturally protected against quantum corrections if they are  pseudo Nambu-Goldstone bosons (pNGBs) of a spontaneously broken global symmetry. In this case, the mass of such particles can be naturally lighter than the scale at which the corresponding symmetry is broken.

A paradigmatic realization of such a possibility is provided by the QCD axion~\cite{Peccei:1977hh, Peccei:1977ur, Wilczek:1977pj, Weinberg:1977ma}, which offers an elegant dynamical solution to the strong CP problem. Inspired by this success, a variety of  light scalar particles have been proposed in connection with other small SM parameters, such as the electroweak scale~\cite{Graham:2015cka}, the flavour puzzle~\cite{Wilczek:1982rv, Calibbi:2016hwq, Ema:2016ops} or neutrino masses~\cite{Chikashige:1980ui, Schechter:1981cv}.
In addition to these, other representative spin-0 particles are given, for instance, by light dilatons~\cite{Salam:1969bwb, Ellis:1970yd, Goldberger:2007zk},  moduli fields in  superstring compactifications~\cite{Cicoli:2013ana} or  Randall-Sundrum scenarios~\cite{Randall:1999ee, Goldberger:1999uk, Goldberger:1999un}, and generic axion-like particles (ALPs) that generalize the couplings of the QCD axion. Such light scalars may be motivated to explain anomalous stellar cooling~\cite{Giannotti:2015dwa, Giannotti:2017hny, Saikawa:2019lng, Badziak:2021apn, DiLuzio:2021ysg}, the abundance of dark matter (DM) in the universe via  e.g., misalignment~\cite{Abbott:1982af, Dine:1982ah, Preskill:1982cy} or freeze-in~\cite{Hall:2009bx}, or, if there are CP-violating interactions with SM fields, the observed dominance of matter over antimatter in the universe \cite{DiLuzio:2020oah, Harigaya:2023bmp}.

Assessing the impact of new light (pseudo-) scalars in these scenarios often requires accurate predictions of their couplings with SM fields at low energies, in terms of the fundamental parameters appearing in specific UV models~\cite{DiLuzio:2020wdo}.
Of particular importance with this respect is the coupling to two photons. Indeed, many experimental searches rely entirely on the interactions of light scalars with the electromagnetic fields~\cite{Jaeckel:2010ni, Graham:2015ouw, Irastorza:2018dyq}, be them axion haloscopes~\cite{Bradley:2003kg, ADMX:2019uok, MADMAX:2019pub} or helioscopes~\cite{IAXO:2019mpb, IAXO:2020wwp, CAST:2024eil},  light-shining-through-walls experiments \cite{Redondo:2010dp, ALPSII:2025eri}, particle colliders~\cite{Mimasu:2014nea, Jaeckel:2015jla, Bauer:2017ris, Bauer:2018uxu} or beam-dump experiments~\cite{Dobrich:2019dxc, NA64:2020qwq, Tammaro:2025zso}. For sufficiently light scalar masses this coupling is severely constrained by astrophysics, see Ref.~\cite{Carenza:2024ehj, Caputo:2024oqc} for recent reviews. Finally, the lifetime of scalars lighter than an MeV is largely set by their coupling to photons\footnote{This is true as long as couplings to neutrinos can be neglected. A direct coupling to neutrinos often features some form of mass suppression due to the chirality-flipping nature of the coupling of any scalar to a pair of on-shell fermions. Loop-induced effects can be generated, but are expected to be suppressed with respect to those to photons by inverse powers of the $W$-boson mass.}, which is crucial in determining their viability as DM candidates. Even if the lifetime exceeds the age of the universe, stringent constraints on sub-MeV DM decays to photons arise from CMB observations~\cite{Slatyer:2016qyl, Bolliet:2020ofj} and X-ray and
low energy $\gamma$-ray telescopes~\cite{Horiuchi:2013noa, Foster:2022ajl,Roach:2022lgo, Laha:2020ivk}, see Ref.~\cite{Panci:2022wlc} for a  summary. 

In the literature, the coupling of light scalars and pseudoscalars to photons at low energies has been extensively studied for the case of CP-preserving couplings, see e.g.~\cite{ Leutwyler:1989tn, GrillidiCortona:2015jxo, Fradette:2018hhl, Lu:2020rhp, Bauer:2020jbp, Ertas:2020xcc, DallaValleGarcia:2023xhh, Aghaie:2024jkj, Bai:2024lpq, Flambaum:2024zyt, Bai:2025fvl, Delaunay:2025lhl, Ovchynnikov:2025gpx}. The purpose of this letter is  to clarify some subtleties related to the computation of the di-photon decay amplitude, and to provide a solid and accurate prediction for the associated decay rate for a light, sub-GeV spinless particle. This calculation will be carried out in terms of the Wilson coefficients of a generic effective field theory (EFT) extending the SM at the electroweak scale, which we set up in Section~\ref{sec:EFT}. Here we also allow for CP-violating interactions, filling an existing gap  in the literature. In Section~\ref{sec:decayrate} we compute the decay rate of light spinless particles to photons, and match our results to various  scenarios in the literature, such as the Higgs Portal, the Dilaton, an axion-like particle and the CPon in Section~\ref{sec:models}. Here we compare to previous calculations, and report various discrepancies with the existing literature. We summarize our conclusions in Section~\ref{sec:conclusions}. Details of the chosen operator basis and scalar-meson mixing are referred to Appendix~\ref{sec:basis} and \ref{sec:mixing}, respectively. 

%%%%%%%%%%%%%%%%%%%%%%%%%%%%%%%%%%%%%%%%%%%%%%%%%%%%%%%%%

\section{Effective Field Theory for a light spin-zero particle}
\label{sec:EFT}
The interactions of a light spinless scalar $\phi$ with SM fields at the electroweak scale are captured by the following Lagrangian:
\begin{align}
\label{Lag1}
\mathcal{L}^{\text{dim-5}} = & - c_{B} \,\frac{g_1^2}{16\pi^2} \frac{\phi}{\Lambda}B_{\mu\nu}B^{\mu\nu}- c_{W} \, \frac{g_2^2}{16\pi^2}\,\frac{\phi}{\Lambda}W_{\mu\nu}W^{\mu\nu}-c_{G}\, \frac{g_s^2}{16\pi^2} \frac{\phi}{\Lambda}G_{\mu\nu}G^{\mu\nu}\nonumber \\
&- \tilde{c}_{B} \,\frac{g_1^2}{16\pi^2}\,\frac{\phi}{\Lambda}B_{\mu\nu}\tilde{B}^{\mu\nu}- \tilde{c}_{W} \,\frac{g_2^2}{16\pi^2}\,\frac{\phi}{\Lambda}W_{\mu\nu}\tilde{W}^{\mu\nu}-\tilde{c}_{G}\,\frac{g_s^2}{16\pi^2}\, \frac{\phi}{\Lambda}G_{\mu\nu}\tilde{G}^{\mu\nu} \nonumber \\
& - \frac{\phi}{\Lambda} \left(\bar{q}_L\,\mathcal{Y}_u\tilde{H}\,u_R+ \bar{q}_L\,\mathcal{Y}_d H\,d_R+ \bar{\ell}_L\,\mathcal{Y}_e H\,e_R\right)  + h.c.\nn\\
&-\beta_1\frac{\phi}{\Lambda}v^2 |H|^2- \beta_2 \frac{\phi}{\Lambda} |H|^4\,,
\end{align}
where $V^{\mu\nu}$, $V=\{B,W_I, G^a\}$, are the field strengths of the $U(1)_Y$, $SU(2)_L$ and $SU(3)_c$ gauge bosons, $\tilde{V}^{\mu\nu} = 1/2 \,\varepsilon^{\mu\nu\rho\sigma} V_{\rho \sigma}$ their duals and $g_1,g_2, g_s$ their respective coupling constants~\footnote{We use the conventions $g_{\mu\nu}={\tt diag}(+1,-1,-1,-1)$,
$\epsilon^{0123}=+1$,
$\gamma_5=i\gamma^0\gamma^1\gamma^2\gamma^3$,
$P_L=\frac{1-\gamma_5}{2}$,
${\tt tr}(\gamma_5\gamma^\mu\gamma^\nu\gamma^\rho\gamma^\sigma)=-4i
\epsilon^{\mu\nu\rho\sigma}$. Under a chiral transformation $f\to f'\equiv e^{-i A\gamma_5}f$, the Lagrangian transforms
as ${\cal L}(f)\to{\cal L}(f')-\frac{\alpha_s}{4\pi}(\sum_{i = {\rm quarks}}  A_{ii}) G_{\mu\nu}\tilde{G}^{\mu\nu}-\frac{\alpha}{4\pi}(\sum_i 2 N_C^i Q_i^2 A_{ii}) F_{\mu\nu}\tilde{F}^{\mu\nu}$.}. The couplings
 $\mathcal{Y}_{\{u,d,e\}}$ are $3 \times 3$  matrices in flavor space. 
 We assume that the dependence of our effective theory on $\phi$ is only through the
dimensionless combination $\phi/\Lambda$, $\Lambda$ denoting a large scale close to the cutoff.
 We expand the theory in powers of $\phi/\Lambda$, neglecting quadratic and higher-order terms. Moreover,
 we adopt a field basis where $\phi$ and $H$ have non-derivative interactions
 and their VEVs read $\langle \phi\rangle=0$, $\langle H\rangle=(0,v/\sqrt{2})^T$ with $v = 246 {\rm \, GeV}$.
 Finally, we adopt a power counting rule where the coupling constants $c_V$ and $\tilde c_V$ are of order $\hbar$,
 while the remaining couplings are all of order $\hbar^0$.
 After electroweak symmetry breaking has occurred, the $SU(3)_\text{c} \times U(1)_{\text{em}}$-invariant terms relevant to the decay of $\phi$ into two photons read in the fermion mass basis~\cite{DiLuzio:2020oah}:
 \begin{align}
\label{eq:Xi_Lag_v1}
\mathcal{L}^{\phi} &=  -\frac{\phi}{\Lambda} \,\bar f \, i y \,\gamma_5 f  - \,\tilde{c}_\gamma\, \frac{\alpha}{4\pi}\,\frac{\phi}{\Lambda}\,  F_{\mu\nu}\tilde{F}^{\mu\nu}  - \tilde c_G\,\frac{\alpha_s}{4\pi} \frac{\phi}{\Lambda}\,   G_{\mu\nu}\tilde{G}^{\mu\nu}\,\nonumber \\
& -\frac{\phi}{\Lambda} \bar f x f -c_\gamma\, \frac{\alpha}{4\pi} \frac{\phi}{\Lambda}\, F_{\mu\nu}F^{\mu\nu}  -c_G\,\frac{\alpha_s}{4\pi}\frac{\phi}{\Lambda}\,  G_{\mu\nu}G^{\mu\nu}-2\xi  m_W^2 \frac{\phi}{\Lambda} W_\mu^+ W^{\mu-}\,,
\end{align}
where
\begin{align}
c_\gamma&=c_B +c_W\, , &\tilde c_\gamma&=\tilde c_B +\tilde c_W \, , \nn\\
x&\equiv  \frac{v}{2 \sqrt{2}}\left(V^\dagger \mathcal{Y}  U +U^\dagger \mathcal{Y}^\dagger V\right)
-\xi m\, , &i y&\equiv  \frac{v}{2 \sqrt{2}}\left(V^\dagger \mathcal{Y}  U -U^\dagger \mathcal{Y}^\dagger V\right)\,.
\end{align}
We use a compact notation where $f$ collects all charged fermions of the SM: $f=(u, d,e)^T$ with generation indices
understood. Similarly we define $\mathcal{Y}=\mathcal{Y}_u\oplus \mathcal{Y}_d\oplus \mathcal{Y}_e$.
Denoting by $Y=Y_u\oplus Y_d\oplus Y_e$ the matrices of Yukawa couplings in the SM: $-(h+v)/\sqrt{2}\bar f_L Y f_R+h.c.$, $V$ and $U$ are unitary matrices such that $m=V^\dagger Y U v/\sqrt{2}$ is diagonal and positive definite.
The matrices  $x$ and $y$ are hermitian.
The parameter $\xi=(\beta_1+\beta_2)/2\lambda$ describes the mixing between $\phi$ and the Higgs. We have neglected
operators proportional to $c_{B,W}$ and $\tilde c_{B,W}$ others than those in Eq.~(\ref{eq:Xi_Lag_v1}),
since their contribution to the $\phi\to\gamma\gamma$ amplitude is of order $\hbar^2$, beyond our level of accuracy.
If there are non-vanishing couplings in both lines of Eq.~(\ref{eq:Xi_Lag_v1}), CP is violated.

As we are interested in discussing the low-energy phenomenology of a spinless state, $m_\phi \ll 1\text{ GeV}$, we need to evolve our Lagrangian from the electroweak scale down to those energies that are typical of the processes we want to probe ($m_\phi$ in the case of the two-photon decay).
In order to do so, we need to take into account both the matching of our Lagrangian on different EFT descriptions as the different mass threshold scales are crossed, and the running of the Wilson coefficients between two subsequent mass scales. 
The running of the Wilson coefficients of the theory from the electroweak scale below has been dealt with in several studies, to which we refer for results \cite{Bauer:2020jbp, Chala:2020wvs, Bonilla:2021ufe, DasBakshi:2023lca, DiLuzio:2023lmd, Bresciani:2024shu, Misiak:2025xzq, Fonseca:2025zjb, Aebischer:2025zxg}.
Regarding the matching procedure, one needs first to integrate out the heavy  fields $t, b, c, \tau$ and $W$~\cite{Bauer:2020jbp, Levati:2024pvl}. Once this is done, one is left with a version of the Lagrangian in Eq.~\eqref{eq:Xi_Lag_v1} that is restricted to interactions of $\phi$ with photons, light leptons, light quarks and gluons. With standard techniques, the interactions among $\phi$ and gluons in the effective Lagrangian can be replaced by other interactions. 
The Lagrangian resulting from these manipulations then reads
\begin{align}
\label{eq:Xil_Lag_QCD_v2}
\mathcal{L}^{\phi} &= -C_\gamma\, \frac{\alpha}{4\pi}\frac{\phi}{\Lambda} F_{\mu\nu}F^{\mu\nu} 
- \tilde{C}_\gamma\, \frac{\alpha}{4\pi}\,\frac{\phi}{\Lambda}\,  F_{\mu\nu}\tilde{F}^{\mu\nu}\,,\nonumber \\&\quad -\frac{\phi}{\Lambda} \bar{\ell} \,x_\ell\, \ell + \frac{\partial_\mu \phi}{\Lambda} \bar{\ell} \,\gamma^\mu \gamma_5 C_\ell\,\ell \,,\nonumber \\
& \quad -\frac{\phi}{\Lambda}\bar{q} \,(X_q+iY_q\gamma_5)\,q \nonumber \\
&\quad+ \text{ tr}(C_q'\lambda_a)\frac{\partial_\mu \phi}{\Lambda} \bar{q}\, \frac{\lambda_a}{2} \gamma^\mu \gamma_5 \,q  +\frac13\text{tr }(C_q')\,\frac{\partial_\mu \phi}{\Lambda} \bar{q}\, \gamma^\mu \gamma_5 \,q\nonumber \\
&\quad+ \kappa\, \frac{\phi}{\Lambda}(\theta^\mu_\mu|_\text{QCD}+\theta^\mu_\mu|_\gamma)\,,
\end{align}

where $\lambda_a$ are the Gell-Mann matrices normalized to $\mathrm{tr}(\lambda_a\lambda_b)=2\delta_{ab}$ and
\begin{align}
\label{cs}
C_\gamma &= c_\gamma - \frac{2}{3}\sum_{i=t,c,b,\tau} N_C^i Q_i^2\,\hat x_i - \frac{\kappa}{2} \beta^{0,\text{quarks}}_{\text{QED}}-\frac72\xi\,,\nonumber \\
\tilde{C}_\gamma&= \tilde c_\gamma +\sum_{i=\mathrm{all}} N_C^i Q_i^2\,\hat y_i  - 4  \tilde{C}_G \sum_{i=u,d,s}N_C^i Q_i^2 Q_{Aii} \,,\nonumber \\
C_G &= c_G - \frac{1}{3}\sum_{i=t,c,b}\,\hat x_i, \nonumber \\
\tilde{C}_G &= \tilde c_G + \frac{1}{2}\sum_{i=\mathrm{quarks}}\hat y_i\,, \nonumber \\
X_{q} &= x_q + \kappa m_q\, , ~~~~~~~~~~~~~~Y_q=2\tilde C_G\{Q_A,m_q\}\,, \nonumber \\
C'_q &= C_q - 2 \tilde{C}_G Q_A\,,~~~~~~~~\kappa = \frac{2 C_G}{\beta_\text{QCD}^0}\nonumber \\
\hat x_i&\equiv\frac{x_{ii}}{m_{i}}\,,~~~~~~~\hat y_i\equiv \frac{y_{ii}}{m_{i}}\,.
\end{align}
Here $Q_i$ and $N_C^i$ are the electric charge and the number of colors of the fermion $f_i$; $Q_A$ is an arbitrary hermitian matrix acting on the space $q=(u,d,s)$ and satisfying $\mathrm{tr}( Q_A)=1/2$~\cite{Georgi:1986df}; $\beta_\text{QCD}^0$ and $\beta^{0,\text{quarks}}_{\text{QED}}$ are the coefficients of one-loop beta functions:
\begin{align}
\beta_\text{QCD}^0=11-\frac23 n_f=9\,,~~~~~~~~~~~~\beta^{0,\text{quarks}}_{\text{QED}}=-\frac43 \sum_{i=u,d,s}N_C^i Q_i^2=-\frac83\,,\nn
\end{align}
while $m_q$, $ x_q, x_\ell,C_q, C_\ell$ represent obvious restrictions to the low-energy degrees of freedom $u,d,s,e,\mu$ of the matrices $m$, $x$ and $C_{ij} = y_{ij}/(m_i+m_j)$. Finally, $\theta^\mu_\mu|_\text{QCD}$ and $\theta^\mu_\mu|_\gamma$ are the traces of the energy-momentum tensor, restricted to the QCD and pure photon contributions, respectively. 
They appear when the operator $\phi GG$ is removed by making use of the trace anomaly equation \cite{Leutwyler:1989tn}. The photon contribution, evaluated in $d$ spacetime dimension, reads
\begin{align}
\theta^\mu_\mu|_\gamma=\frac{d-4}{4}F_{\mu\nu}F^{\mu\nu}\,,\nn
\end{align}
an evanescent term needed for the consistency of the $\phi$ decay amplitude. 
In eqs. (\ref{eq:Xil_Lag_QCD_v2}, \ref{cs}), we have kept the leading terms of the expansion in powers of $1/M$, $M$ representing the mass of an heavy state such as $t$, $b$, $c$, $\tau$ and $W$. In particular we have neglected contributions to the effective photon couplings that are suppressed by $m_\phi^2/M^2$.

We are mainly interested in a mass range for $\phi$  where the dominant decay channel is into two photons
(or two neutrinos). Since our effective theory should be applied
at energy scales smaller than the QCD confinement scale, the interactions with quarks and gluons have to be treated by making use of non perturbative techniques. In this relation, chiral perturbation theory ($\chi$pt) represents a suitable framework: interactions with quarks and gluons can be traded for interactions with those degrees of freedom that are relevant at low energies, light pseudoscalar mesons in our case. The procedure has been discussed in detail in \cite{Georgi:1986df, Leutwyler:1989xj, Bauer:2020jbp, Arina:2021nqi, DiLuzio:2023cuk}, and here we simply summarize their results. First of all, $\phi$ interactions with quarks are introduced in the chiral Lagrangian as external sources. The chiral counterpart to any quark-containing operator in the Lagrangian in Eq.~\eqref{eq:Xi_Lag_v1} can be found by making use of the low-energy quark-hadron duality. Such a procedure cannot be applied to gluon-containing operators, which have been removed from the Lagrangian of Eq.~(\ref{eq:Xil_Lag_QCD_v2}) in favor of other operators. 
The chiral counterpart to the Lagrangian in Eq.~\eqref{eq:Xil_Lag_QCD_v2}, valid up to order $\mathcal{O}(p^2)$ in the chiral expansion and up to order $\mathcal{O}(\Lambda^{-2})$ in the $\phi$EFT expansion, is found to be
\begin{align}
\mathcal{L}_{\phi}^{\chi\text{pt}} =\mathcal{L}_{\phi PC}^{\chi\text{pt}} +\mathcal{L}_{\phi PV}^{\chi\text{pt}}\,,\nn
\end{align}
where, by assigning conventionally a positive parity to $\phi$, we have separated parity-conserving (PC) and parity-violating (PV) contributions. These read
\begin{align} 
\mathcal{L}_{\phi PC}^{\chi\text{pt}}=
&-C_\gamma\, \frac{\alpha}{4\pi}\frac{\phi}{\Lambda} F_{\mu\nu}F^{\mu\nu}-\frac{\phi}{\Lambda} \bar{\ell} \,x_\ell\, \ell \nn\\
&+ \frac{f_\pi^2}{2} B_0 \frac{\phi}{\Lambda} \text{ tr} \left[X_q(\Sigma + \Sigma^\dagger)\right] \nonumber \\
&- \kappa \frac{f_\pi^2}{2}\frac{\phi}{\Lambda} \left\{\text{tr }(D_\mu\Sigma D^\mu\Sigma^\dagger)
+4 B_0 \text{ tr}\left[m_q(\Sigma+ \Sigma^\dagger)\right]\right\}+2\kappa\frac{\phi}{\Lambda}M_0^2\eta_0^2\nn\\
& + \kappa\frac{d-4}{4}\frac{\phi}{\Lambda}F_{\mu\nu}F^{\mu\nu}\,,
\label{PC}\\
&\nn\\
\mathcal{L}_{\phi PV}^{\chi\text{pt}}=
&- \tilde{C}_\gamma\, \frac{\alpha}{4\pi}\,\frac{\phi}{\Lambda}\,  F_{\mu\nu}\tilde{F}^{\mu\nu}+ \frac{\partial_\mu \phi}{\Lambda} \bar{\ell} \,\gamma^\mu \gamma_5 C_\ell\,\ell \,\nonumber\\
&+\frac{\partial_\mu \phi}{\Lambda} \left[ \text{ tr } (C_q'\lambda_a) j_A^{\mu,a}
+ \frac13 \text{tr }(C_q')j_A^\mu\right]\nonumber\\
&+ i\,  \tilde C_G  f_\pi^2 B_0 \frac{\phi}{\Lambda} \text{ tr} \left[\{m_q, Q_A\}(\Sigma^\dagger - \Sigma)\right] 
\label{PV}\,,
\end{align}
where $f_\pi = 92.4\pm 0.3 \text{ MeV}$, $B_0 = m_{\pi^\pm}^2/(m_u+m_d)$, $M_0$ is the $\eta_0$ mass in the chiral limit and
\begin{align}
\Sigma&=\exp \left[ i \sqrt{2} \Phi/ f_\pi\right]\,,~~~~~
\Phi=
\left(
\begin{array}{ccc}
\frac{1}{\sqrt{2}}\pi^0+\frac{1}{\sqrt{6}}\eta_8&\pi^+&K^+\\
\pi^-&-\frac{1}{\sqrt{2}}\pi^0+\frac{1}{\sqrt{6}}\eta_8&K^0\\
K^-&\bar K_0&-\frac{2}{\sqrt{6}}\eta_8
\end{array}
\right)+\frac{1}{\sqrt{3}}\eta_0\mathbbm{1}\,,\nn\\
D_\mu \Sigma& =\partial_\mu \Sigma + i e A_\mu \left[Q_q, \Sigma\right], \nonumber \\
j_{A,a}^\mu &= i \frac{f_\pi^2}{2} \text{ tr } \left[\frac{\lambda_a}{2}(D^\mu \Sigma^\dagger) \Sigma - \frac{\lambda_a}{2}(D^\mu \Sigma)  \Sigma^\dagger\right]\,,~~~~~
j_{A}^\mu = i \frac{f_\pi^2}{2} \text{ tr } \left[(D^\mu \Sigma^\dagger) \Sigma -(D^\mu \Sigma)  \Sigma^\dagger\right]\,.\nn\\
\end{align}

\noindent
Under the chiral group SU(3)$_L\times$SU(3)$_R$, $\Sigma(x)$ transforms as $\Sigma(x)\to L \Sigma(x) R^\dagger$.
The last two terms of $\mathcal{L}_{\phi PV}^{\chi\text{pt}}$ induce a mixing among $\phi$ and the light pseudoscalar mesons at the level of kinetic and mass terms. This mixing generates a coupling of $\phi$ to photons, inherited from the Wess-Zumino-Witten (WZW) term accounting for the anomalies 
of the global currents~\cite{Wess:1971yu, Witten:1983tw}. In particular, the chiral Lagrangian  contains the operator
\be
{\cal L} \supset -\frac{\alpha}{4\pi f_\pi} F_{\mu\nu} \tilde F^{\mu\nu}  \sum_{P = \pi^0, \eta, \eta^\prime} C_P P \,,
\ee
with $C_\pi  =  1, C_\eta = 4/3 \sqrt{2/3}, C_{\eta^\prime} = 7/(3 \sqrt 3)$, and where we have fixed the $\eta-\eta^\prime$ mixing angle $\alpha_{\eta \eta^\prime}$ to about $-19^\circ$, see Appendix~\ref{sec:mixing} for details. Expressing the mesons $P$ in terms of the mass eigenstate $\phi$ as $P = \theta_{P\phi} \phi + \hdots$,  an effective $\phi$-gamma-gamma interaction $-\delta \tilde C_\gamma\frac{\alpha}{4\pi}\frac{\phi}{\Lambda}F_{\mu\nu} \tilde F^{\mu\nu}$ is induced with \begin{align}
\delta\tilde C_\gamma & = \frac{\Lambda}{f_\pi} \sum_{P = \pi^0, \eta, \eta^\prime} C_P \, \theta_{P \phi}  \,. 
\end{align}
We have calculated the mixing angles in Appendix~\ref{sec:mixing}, systematically expanding either in the isospin-breaking parameter $\delta = (m_d-m_u)/(m_d+m_u)$ or in the small ratio $m_{\phi,\pi}^2/m_\eta^2$, for  general matrices $Q_A$ with ${\rm tr \, Q_A = 1/2}$. 

In conclusion, for the purpose of calculating the decay amplitude of $\phi$ into two photons at the desired level of approximation, we will use the Lagrangian $\mathcal{L}_{\phi}^{\chi\text{pt}} =\mathcal{L}_{\phi PC}^{\chi\text{pt}} +\mathcal{L}_{\phi PV}^{\chi\text{pt}}$, where $\mathcal{L}_{\phi PC}^{\chi\text{pt}}$ is as in Eq.~(\ref{PC}), while 
the relevant part of $\mathcal{L}_{\phi PV}^{\chi\text{pt}}$ reads
\begin{align}
\mathcal{L}_{\phi PV}^{\chi\text{pt}}=
- \tilde{C}'_\gamma\, \frac{\alpha}{4\pi}\,\frac{\phi}{\Lambda}\,  F_{\mu\nu}\tilde{F}^{\mu\nu}+ \frac{\partial_\mu \phi}{\Lambda} \bar{\ell} \,\gamma^\mu \gamma_5 C_\ell\,\ell \,,
\end{align}
where $\tilde{C}'_\gamma=\tilde{C}_\gamma+\delta \tilde{C}_\gamma$, which reads
\begin{align}
\label{tildeCgamma}
\tilde{C}'_\gamma &= \tilde c_\gamma +\sum_{i=\mathrm{all}} N_C^i Q_i^2\,\hat y_i 
+ \frac{  \hat y_{u} - \hat y_{d}}{2}\frac{m_\phi^2}{m_\pi^2 - m_\phi^2}  \nn \\
& + \frac{m_\phi^2}{24m_\eta^2} \left[ 13 \hat y_{u} + 13 \hat y_{d} - 6 \hat y_{s}  -  \delta\, \frac{m_\pi^2}{m_\pi^2 - m_\phi^2} \left( 4 \hat y_{d} + 6 \hat y_{s} - 22 \hat y_{u}  \right) \right] \nn \\ 
&+\tilde{C}_G\left[- \frac{5}{3} -  \delta\,    \frac{m_\pi^2}{m_\pi^2 - m_\phi^2} + \frac{13 m_\pi^2}{12m_\eta^2}  + \delta\, ( 9 - 13 \delta - 9 \delta^2  ) \frac{ m_\pi^4}{12m_\eta^2 (m_\pi^2 - m_\phi^2 )}\right]   
\nn\\
& - \frac{m_\phi^2}{24m_\eta^2} \tilde{C}_G\left[  26   +  18\,\delta\, \frac{ m_\pi^2}{ m_\pi^2 - m_\phi^2} \right]  \, ,
\end{align}
where we have calculated the $\delta \tilde{C}_\gamma$ contribution using the general results in Appendix~\ref{sec:mixing}, employing the result of the $m_{\phi, \pi}^2/m_\eta^2$ expansion in  Eq.~(\ref{RZ}),  fixing $\alpha_{\eta \eta^\prime} \approx -19^\circ$.  One can explicitly verify that the $Q_A$ dependence cancels out in $\tilde{C}_\gamma^\prime$.

%%%%%%%%%%%%%%%%%%%%%%%%%%%%%%%%%%%%%%%%%%%%%%%%%%%%%%%%
\section{Decay rate}
\label{sec:decayrate}
\begin{figure}[h!]
\centering
    \includegraphics[width=\linewidth]{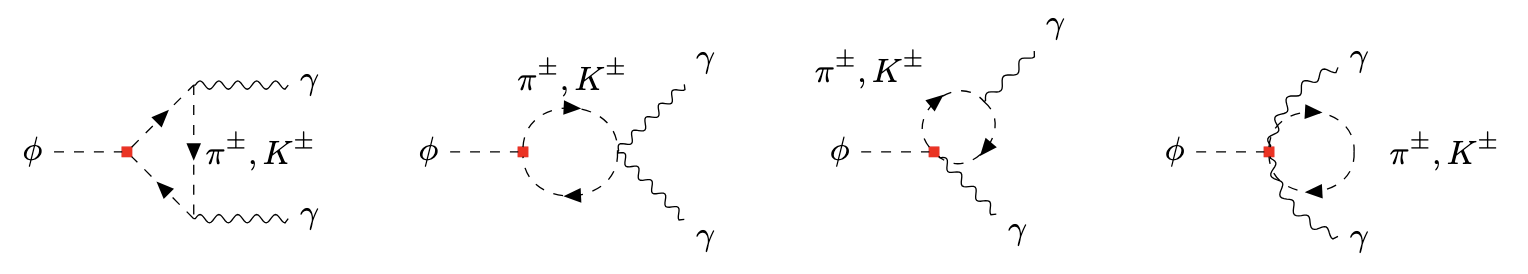}\\[0.3 cm]
        \includegraphics[width=0.5\linewidth]{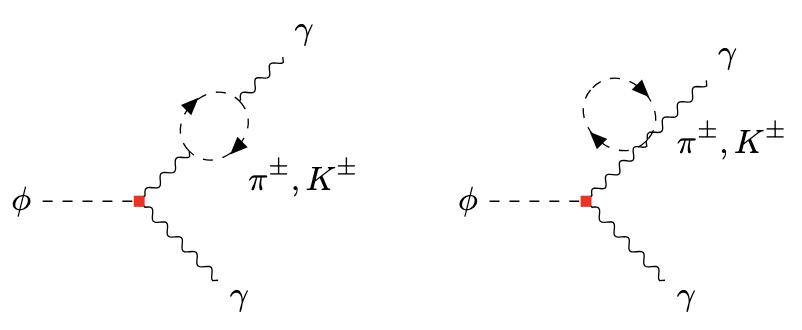}
    \caption{One-loop diagrams contributing to $C_\gamma^\text{eff}(m_\phi)$.
    The tree-level diagram  proportional to $d-4$ vanishes in $d=4$ dimensions, but produces a finite contribution when combined with the external-leg divergences induced by the last two diagrams.}
    \label{fig:One_Loop_Mesonic_Vertex}
\end{figure}

%%%%%%%%%%%%%%%%%%%%%%%%%%%%%%%%%%%%%%%%%%%%%%%%%%%%%%%%
The decay rate of $\phi$ into two photons can be expressed as
\begin{equation}
\label{decayrate}
\Gamma(\phi \to \gamma \gamma) = \frac{\alpha^2}{64\pi^3}\,\frac{m_\phi^3}{\Lambda^2}\,(|C_\gamma^\text{eff}(m_\phi)|^2+|\tilde{C}_\gamma^\text{eff}(m_\phi)|^2)\,,
\end{equation}
\begin{comment}
where $C_\gamma^\text{eff}$ and $\tilde{C}_\gamma^\text{eff}$ are the coefficients of the Lagrangian
\begin{equation}
\mathcal{L}^\text{dim-5}_{\phi \to \gamma \gamma} = - \frac{\alpha}{4\pi} \frac{\phi}{\Lambda}\left( C_\gamma^\text{eff}  F_{\mu\nu}F^{\mu\nu} +\tilde{C}_\gamma^\text{eff}  F_{\mu\nu} \tilde{F}^{\mu\nu} \right)\,, 
\end{equation}
\end{comment}
obtained by evaluating the two-photon decay amplitude from $\mathcal{L}_{\phi}^{\chi\text{pt}}$.
We provide explicit expressions for the amplitudes $C_\gamma^\text{eff}(m_\phi)$ and $\tilde{C}_\gamma^\text{eff}(m_\phi)$ up to one-loop level.  At this order, $C_\gamma^\text{eff}(m_\phi)$ and $\tilde{C}_\gamma^\text{eff}(m_\phi)$ depend separately on $\mathcal{L}_{\phi PC}^{\chi\text{pt}}$ and $\mathcal{L}_{\phi PV}^{\chi\text{pt}}$, respectively. We find
\begin{align}
\label{rPC}
C_\gamma^\text{eff}(m_\phi)&=C_\gamma
-\sum_{i=e,\mu}\hat{x}_i
J_{1/2}^{PC}\left(\frac{m_\phi}{2 m_i} \right)
- \kappa\sum_{M=\pi,K}\left[\left(\frac14+\frac{m_\phi^2}{4 m_M^2}\right)J_0\left(\frac{m_\phi}{2 m_M} \right)+\frac16\right]\nn\\
&+\frac14\frac{x_{uu}+x_{dd}}{m_u+m_d}J_0\left(\frac{m_\phi}{2 m_\pi} \right)+\frac14\frac{x_{uu}+x_{ss}}{m_u+m_s}J_0\left(\frac{m_\phi}{2 m_K} \right) \, , \nn\\[0.3 cm]
\tilde{C}_\gamma^\text{eff}(m_\phi)&=\tilde{C}'_\gamma-\sum_{i=e,\mu}\hat{y}_i J_{1/2}^{PV}\left(\frac{m_\phi}{2 m_i} \right)
\,,
\end{align}
where 
\begin{align}
J_{1/2}^{PC}(z)&=\frac{1}{z^2}\left[1+(\arcsin z)^2\right]-\frac{1}{z^4}(\arcsin z)^2 \xrightarrow[z \ll 1]{} \frac{2}{3} + \frac{7}{45} z^2 + \hdots  \nn \\
J_0(z)&=\frac{1}{z^2}-\frac{1}{z^4}(\arcsin z)^2 \xrightarrow[z \ll 1]{} - \frac{1}{3} - \frac{8}{45} z^2 + \hdots \nn  \\
J_{1/2}^{PV}(z)&=1-\frac{1}{z^2}(\arcsin z)^2 \xrightarrow[z \ll 1]{} -  \frac{1}{3} z^2 + \hdots
\end{align}
The terms proportional to $\kappa$ are sourced by the operator $\phi\, G_{\mu\nu} G^{\mu\nu}$ and are in agreement with the results of Ref.~\cite{Leutwyler:1989tn}. They are obtained by summing the diagrams in Fig.~\ref{fig:One_Loop_Mesonic_Vertex}.

For $m_\phi$ smaller that about $1 \MeV$, the above coefficients are real and the $\phi\gamma\gamma$ interaction
can be described by the effective Lagrangian
\begin{align}
{\cal L}_{\phi\gamma\gamma}
&=-\frac{\alpha}{4\pi}\frac{\phi}{\Lambda}\left[
C_\gamma^\text{eff}(m_\phi)F_{\mu\nu}F^{\mu\nu}+\tilde{C}_\gamma^\text{eff}(m_\phi)F_{\mu\nu}\tilde{F}^{\mu\nu}
\right]\,.
\end{align}
It is instructive to examine the approximate expressions of $C_\gamma^\text{eff}(m_\phi)$ and $\tilde{C}_\gamma^\text{eff}(m_\phi)$ when $m_\phi$ is small. In this regime, we can expand the functions $J_0(z)$ and $J_{1/2}^{PC,PV}(z)$ around $z=0$. By keeping the first two orders in the series expansion and
 recalling the expression of $C_\gamma$ and $\tilde{C}'_\gamma$ in Eq.~\eqref{cs} and \eqref{tildeCgamma}, we obtain:
\begin{align}
\label{ceff}
C_\gamma^\text{eff}(m_\phi)&=c_\gamma
-\frac23\sum_{i=e,\mu,\tau,c,b,t}N_C^i Q_i^2\hat{x}_i+\frac{7}{6}\kappa-\frac72\xi-\frac{1}{12}\left(\frac{x_{uu}+x_{dd}}{m_u+m_d}+\frac{x_{uu}+x_{ss}}{m_u+m_s}\right)\\
&-\frac{7}{180}\sum_{i=e,\mu}\hat{x}_i
\frac{m_\phi^2}{m_i^2}
+\frac{17 \kappa}{180}\left(\frac{m_\phi^2}{m_\pi^2}+\frac{m_\phi^2}{m_K^2}\right)
-\frac{1}{90}\left(\frac{x_{uu}+x_{dd}}{m_u+m_d}\frac{m_\phi^2}{m_\pi^2}+\frac{x_{uu}+x_{ss}}{m_u+m_s}\frac{m_\phi^2}{m_K^2}\right)+ \hdots \nn\\[0.3 cm]
\tilde{C}_\gamma^\text{eff}(m_\phi)&=
\tilde c_\gamma +\sum_{i=\mathrm{all}} N_C^i Q_i^2\,\hat y_i -\tilde{C}_G\left(\frac{5}{3} +  \delta\,    \frac{m_\pi^2}{m_\pi^2 - m_\phi^2}\right)\nn\\
&+\tilde{C}_G\left[ \frac{13 m_\pi^2}{12m_\eta^2}  + \delta\, ( 9 - 13 \delta - 9 \delta^2  ) \frac{ m_\pi^4}{12m_\eta^2 (m_\pi^2 - m_\phi^2 )}\right]
\\
& + \frac{  \hat y_{u} - \hat y_{d}}{2}\frac{m_\phi^2}{m_\pi^2 - m_\phi^2} - \frac{m_\phi^2}{24m_\eta^2} \tilde{C}_G\left[  26   +  18\,\delta\, \frac{ m_\pi^2}{ m_\pi^2 - m_\phi^2} \right]+\frac{1}{12}\sum_{i=e,\mu}\hat{y}_i\frac{m_\phi^2}{m_i^2}\nn\\
& + \frac{m_\phi^2}{24m_\eta^2} \left[ 13 \hat y_{u} + 13 \hat y_{d} - 6 \hat y_{s}  -  \delta\, \frac{m_\pi^2}{m_\pi^2 - m_\phi^2} \left( 4 \hat y_{d} + 6 \hat y_{s} - 22 \hat y_{u}  \right) \right]+ \hdots  \nn
\end{align}
When $\phi$ is lighter than about 1 MeV, barring cancellation of  the leading-order terms, the decay rate is dominated
by the values of these coefficients evaluated at $m_\phi^2=0$. Making use of the parametrization
\begin{align}
\label{C0}
C_\gamma^\text{eff}(0)&=\chi_\gamma\, c_\gamma+\chi_G\, c_G+\chi_\xi\, \xi+\sum_{i=\mathrm{all}} \chi_i\, \hat x_{i}\,,\nn\\
\tilde C_\gamma^\text{eff}(0)&=\tilde\chi_\gamma\, \tilde c_\gamma+\tilde\chi_G\, \tilde c_G+\sum_{i=\mathrm{all}} \tilde\chi_i\, \hat y_{i}\,,
\end{align}
and chosing as inputs $m_u=2.2$ MeV, $m_d=4.7$ MeV, $m_s=93.5$ MeV, $\delta=0.36$, $m_{\pi^0}=135$ MeV, $m_\eta=547.9$ MeV~\cite{ParticleDataGroup:2024cfk},  we get the numerical estimate 
displayed in Table \ref{coeff}.
\begin{table}[h!]
\centering
\resizebox{1.\textwidth}{!}{
\begin{tabular}{|c|c|c|c|c|c|c|c|c|c|c|c|} 
\hline
$\chi_\gamma$&$\chi_G$&$\chi_e$&$\chi_\mu$&$\chi_\tau$&$\chi_u$&$\chi_d$&$\chi_s$&$\chi_c$&$\chi_b$&$\chi_t$&$\chi_\xi$\\
\hline
$1$&$0.26$&$-0.67$&$-0.67$&$-0.67$&$-0.03$&$-0.06$&$-0.08$&$-0.98$&$-0.31$&$-0.98$&$-3.5$\\
\hline
\hline
$\tilde\chi_\gamma$&$\tilde\chi_G$&$\tilde\chi_e$&$\tilde\chi_\mu$&$\tilde\chi_\tau$&$\tilde\chi_u$&$\tilde\chi_d$&$\tilde\chi_s$&$\tilde\chi_c$&$\tilde\chi_b$&$\tilde\chi_t$&\\
\hline
$1$&$-1.96$&$1$&$1$&$1$&$0.35$&$-0.65$&$-0.65$&$0.35$&$-0.65$&$0.35$&\\
\hline
\end{tabular}}
\caption{\label{coeff}
Coefficients of $C_\gamma^\text{eff}(0)$ and $\tilde C_\gamma^\text{eff}(0)$, in the parametrization of Eq.~(\ref{C0}).}
\end{table}
When $m_\phi<1$ MeV the correction to Eq.~(\ref{C0}) are very small and are expected to be dominated by the
additional contributions
$\Delta C_\gamma^{\rm eff} \approx-0.04\, \hat x_e m_\phi^2/m_e^2$, $\Delta \tilde{C}_\gamma^{\rm eff} \approx+0.08\, \hat y_e m_\phi^2/m_e^2$.
%%%%%%%%%%%%%%%%%%%%%%%%%%%%%%%%%%%%%%%%%%%%%%%%%%%%%%%%
%%%%%%%%%%%%%%%%%%%%%%%%%%%%%%%%%%%%%%%%%%%%%%%%%%%%%%%%
\section{Matching to specific models}
\label{sec:models}
%%%%%%%%%%%%%%%%%%%%%%%%%%%%%%%%%%%%%%%%%%%%%%%%%%%%%%%%
\begin{itemize}
\item {\bf Light CP-even scalar and Higgs Portal}

If $\phi$ is a genuine scalar particle with diagonal fermion couplings ${\cal L} \supset - \phi/\Lambda \overline{f}_i x_i f_i$, its decay rate into two photons is obtained by setting to zero the 
coupling $\tilde C_\gamma^\text{eff}(m_\phi)$ in Eq.~(\ref{decayrate}) and taking  $C_\gamma^\text{eff}(m_\phi)$ given by the expression in Eq.~(\ref{rPC}) with $c_\gamma = c_G =0, \hat{x}_i = x_i/m_i,  \kappa = - 2/27 (x_t/m_t + x_c/m_c + x_b/m_b) , \xi = 0$. When $m_\phi\ll 1$ MeV,
the effective photon coupling is well approximated by 
\begin{align}
C_\gamma^\text{eff}(0)& = 
-\frac23\sum_{i=e,\mu,\tau,c,b,t}N_C^i Q_i^2 \frac{x_i}{m_i}-\frac{1}{12}\left(\frac{x_{u}+x_{d}}{m_u+m_d}+\frac{x_{u}+x_{s}}{m_u+m_s}\right) -\frac{7}{81}  \sum_{i=t,c,b}  \frac{x_i}{m_i} \, .
\end{align}
This result slightly differs from the one of Ref.~\cite{Flambaum:2024zyt}, where the effective photon-scalar coupling, evaluated to one-loop accuracy, does not include the last contribution from heavy quarks proportional to $\kappa$.

In Higgs Portal models $\phi$ is identified with a light CP-even scalar that couples to the SM particles only via mixing with the Higgs, i.e. only  the last terms in Eq.~\eqref{Lag1} are non-zero. Keeping only the resulting mixing angle $\xi$ in the final expressions (note that in this limit $x = - \xi m$), we obtain
\begin{align}
C_\gamma^\text{eff}(m_\phi)& =   \frac{2 \xi}{3}  \sum_{i=t,c,b,\tau} N_C^i Q_i^2 + \frac{8 \xi}{27}   -\frac{7\xi}{2} 
+ \xi \sum_{i=e,\mu}
J_{1/2}^{PC}\left( \frac{m_\phi}{2 m_i}\right) \nonumber \\
& -  \frac{2 \xi}{9}   \sum_{M=\pi,K}\left[\left(\frac14+\frac{m_\phi^2}{4 m_M^2}\right)J_0\left( \frac{m_\phi}{2 m_M}\right)+\frac16\right] - \frac{\xi}{4} \left( J_0\left( \frac{m_\phi}{2 m_\pi}\right) + J_0\left( \frac{m_\phi}{2 m_K}\right) \right) \nn \\
& = \xi \left( - \frac{11}{18} +  \sum_{i=e,\mu}  J_{1/2}^{PC}\left( \frac{m_\phi}{2 m_i}\right) -  \frac{1}{18}   \sum_{M=\pi,K}\left[\left(\frac{11}{2}  +\frac{m_\phi^2}{ m_M^2}\right)J_0\left( \frac{m_\phi}{2 m_M}\right)   \right] \right)  \, , 
\end{align}
in agreement with Ref.~\cite{Leutwyler:1989tn}. For very light scalars with $m_\phi \ll 2 m_e$ ones has
\begin{align}
C_\gamma^\text{eff}(0)&  = \xi \left( - \frac{11}{18} + \frac{4}{3} +  \frac{11}{54}  \right) =  \frac{25}{27} \xi \, , 
\end{align}
in agreement with Ref.~\cite{Fradette:2018hhl}.

\item{\bf Dilaton}

The dilaton, the pseudo-Goldstone boson associated with the spontaneous breaking of scale invariance,
has been the subject of extensive study for many years~\cite{Salam:1969bwb, Ellis:1970yd, Randall:1999ee, Goldberger:1999un, Goldberger:2007zk}. Its phenomenology has been discussed within an EFT framework \cite{Vecchi:2010gj, Chacko:2012sy, Ahmed:2019csf, Appelquist:2022mjb}, and specific mechanisms have been designed to obtain a  sub-GeV light dilaton \cite{Bellazzini:2013fga,Coradeschi:2013gda,Abu-Ajamieh:2017khi,Megias:2015qqh, Agrawal:2016ubh}. Here we consider a simple illustrative example, consisting of the SM degrees of freedom supplemented --- at low energies --- by an extra light dilaton $\sigma$, coupled to those operators that break scale invariance, either at the classical or at the quantum level. 
In particular, we focus on the following dilaton interactions
\begin{align}
\label{eq:Dilaton_SM_int}
\mathcal{L}_{\sigma\text{-SM}}^\text{int} &= \frac{\sigma}{\Lambda} \left(-2 m_W^2 W^-_\mu W^{+\mu}- m_Z^2 Z_\mu Z^\mu + \sum_f m_f\, \bar f f + m_h^2 h^2\right)\nn\\
& - \frac{\sigma}{\Lambda}\left(\frac{\alpha_\text{Y}}{8\pi} b^0_\text{Y} B^{\mu\nu}B_{\mu\nu}+\frac{\alpha_2}{8\pi} b^0_2 W^{a\mu\nu}W^a_{\mu\nu}+ \frac{\alpha_\text{s}}{8\pi} b^0_\text{QCD} G_a^{\mu\nu}G^a_{\mu\nu}\right)\,,
\end{align}
written in the mass eigenstate basis, identifying the scale-invariance breaking scale with $\Lambda$. 
Here $B_{\mu\nu}$ and $W^a_{\mu\nu}$ are the field strengths of the electroweak gauge vector bosons
with gauge couplings $\alpha_\text{Y}=g'^{2}/4\pi$ and $\alpha_2=g^2/4\pi$, respectively.

At the classical level, for the purpose of computing the decay amplitude into two photons, the dilaton couplings coincide with those of a scalar that mixes with the Higgs boson. Matching to the Lagrangian in Eq. (\ref{eq:Xi_Lag_v1}) yields $\xi=1$ and $\hat x_i=-1$.
Beyond the classical term in Eq.~(\ref{eq:Dilaton_SM_int}), the dilaton also couples to the anomalous trace of the energy momentum tensor. The coefficients $b^0_\text{Y}$, $b^0_2$ and $b^0_\text{QCD}$ are the beta-function coefficient of the SM gauge couplings, and we get $c_\gamma=b^0_\text{QED}/2=(b^0_\text{Y}+b^0_2)/2$ and $c_G=b^0_\text{QCD}/2$~\footnote{We distinguish $b^0_\text{QED}$ and $b^0_\text{QCD}$ from $\beta^{0,\text{quarks}}_{\text{QED}}=-\frac83$ and $\beta_\text{QCD}^0=9$, that have been introduced to match QCD and chiral Lagrangians at the GeV scale, with three active flavours.}.

The couplings of the dilaton to massive gauge bosons and to fermions are corrected at the loop level by the corresponding anomalous mass dimension, which is a further source of scale symmetry breaking. 
Moreover, for the dilaton to acquire a mass, explicit  scale-invariance violating terms should be present, which
could alter the couplings in Eq.~(\ref{eq:Dilaton_SM_int}). We assume both these effects to be negligible in a first approximation.

We get:
\begin{align}
\label{dilaton}
C_\gamma^\text{eff}(m_\phi)&=\frac{b^0_\text{QED}}{2}+\frac23  \sum_{i=t,c,b,\tau} N_C^i Q_i^2+\frac{8}{27}\left(\frac{b^0_\text{QCD}}{2}+1\right)-\frac72\nn\\
&+\sum_{i=e,\mu}  J_{1/2}^{PC}\left( \frac{m_\phi}{2 m_i}\right)
-  \frac{2}{9} \left(\frac{b^0_\text{QCD}}{2}+1\right)  \sum_{M=\pi,K}\left[\left(\frac14+\frac{m_\phi^2}{4 m_M^2}\right)J_0\left( \frac{m_\phi}{2 m_M}\right)+\frac16\right]\nn\\
&- \frac14 \left[ J_0\left( \frac{m_\phi}{2 m_\pi}\right)+ J_0\left( \frac{m_\phi}{2 m_K}\right)\right] \,  , \nn\\[0.3 cm]
\tilde{C}_\gamma^\text{eff}(m_\phi)&=0\,.
\end{align}

Assuming $b^0_\text{QED}$ and $b^0_\text{QCD}$ saturated by all SM degrees of freedom, that is $b^0_\text{QED}=-41/6+19/6=-11/3$ and $b^0_\text{QCD}=7$, for a very light dilaton with $m_\sigma \ll 2 m_e$ we get:
\begin{align}
\label{dilaton}
C_\gamma^\text{eff}(m_\sigma)=\frac{7}{180}\left(
\frac{m_\sigma^2}{m_e^2}+\frac{m_\sigma^2}{m_\mu^2}\right)
+\frac{1}{10}\left(\frac{m_\sigma^2}{m_\pi^2}+\frac{m_\sigma^2}{m_K^2}\right)+ \hdots
\end{align}
The decay amplitude of the dilaton into two photons vanishes in the limit $m_\sigma=0$, in agreement with general
properties of the matrix element $\langle 0|\theta_\mu^\mu|\gamma\gamma'\rangle$,
relying on Lorentz invariance, gauge invariance, Bose statistics and energy-momentum conservation~\cite{Leutwyler:1989tn}.

\item {\bf Axion-like Particles}

For axion-like particles $a$ the Lagrangian is usually written in a basis where the ALP has only derivative couplings. As only flavor-diagonal terms contribute to the decay rate, we restrict to such couplings, described by the Lagrangian
\begin{align}
{\cal L} & =  N \frac{a}{\Lambda} \frac{\alpha_s}{4 \pi} G_{\mu \nu} \tilde{G}^{\mu \nu} + E \frac{a}{\Lambda} \frac{\alpha_{\rm em}}{4 \pi} F_{\mu \nu} \tilde{F}^{\mu \nu} + \frac{\partial_\mu a}{2 \Lambda}  \sum_i c_i \overline{f}_i \gamma^\mu \gamma_5  f_i \, ,
\end{align}
In order to match to the CP-odd Lagrangian in the first line of Eq.~\eqref{eq:Xi_Lag_v1} with $\phi = a$, we perform chiral fermion field redefinitions proportional to the parameter $c_i$
\begin{align}
f_i & \to e^{i c_i a(x) /2 \Lambda \gamma_5} f_i  \, , 
\end{align}
which remove the derivative couplings and give (up to higher powers of $a$) 
\begin{align}
\tilde{c}_G & = - N - \frac{1}{2} \sum_{i = \rm quarks} c_i \, ,  &  \tilde{c}_\gamma & = - E -  \sum_{i = \rm all}   N_C^i Q_i^2 c_i  \, , & \hat y_i & = c_i  \, , 
\end{align}
so that $\tilde{C}_G  = - N$, which finally gives $C_\gamma^\text{eff}(m_a) = 0$ and
\begin{align}
\label{alpresult}
\tilde{C}_\gamma^\text{eff}(m_a) &= -E
+ \frac{  c_u - c_d}{2}\frac{m_a^2}{m_\pi^2 - m_a^2}  \nn \\
& + \frac{m_a^2}{24m_\eta^2} \left[ 13 c_u + 13 c_d - 6 c_s  -  \delta\, \frac{m_\pi^2}{m_\pi^2 - m_a^2} \left( 4 c_d + 6 c_s - 22 c_u  \right) \right] \nn \\ 
&-N\left[- \frac{5}{3} -  \delta\,    \frac{m_\pi^2}{m_\pi^2 - m_a^2} + \frac{13 m_\pi^2}{12m_\eta^2}  + \delta\, ( 9 - 13 \delta - 9 \delta^2  ) \frac{ m_\pi^4}{12m_\eta^2 (m_\pi^2 - m_a^2 )}\right]   
\nn\\
& + \frac{m_a^2}{24m_\eta^2} N\left[  26   +  18\,\delta\, \frac{ m_\pi^2}{ m_\pi^2 - m_a^2} \right]-\sum_{i=e,\mu} c_i J_{1/2}^{PV} \left( \frac{m_a}{2 m_i}\right) \, .
\end{align}
This result extends and corrects various results in the literature, see Appendix~\ref{sec:mixing} for details. In the limit $m_\eta \to \infty$ we recover the results of Ref.~\cite{Bauer:2020jbp}. Terms suppressed by $1/m_\eta^2$ have been calculated to first order in $\delta$ in Ref.~\cite{Bai:2025fvl}, but we disagree on terms proportional to $N$ in the isospin limit (see Section~\ref{sec:deltaexpansion}), presumably because of typographical errors. 

In the limit of tiny ALP masses (like the standard QCD axion), $m_a \ll m_e$ we obtain
\begin{align}  
\tilde{C}_\gamma^\text{eff}(0) &= -E
-N\left[- \frac{5}{3} -  \delta  +  (13 + 9 \delta - 13 \delta^2 - 9 \delta^3  ) \frac{ m_\pi^2}{12m_\eta^2 }\right]   \approx  - E  + 1.96 N  \, .
\end{align}
This results captures only partially the complete NLO corrections in the chiral expansion that have been derived in Ref.~\cite{GrillidiCortona:2015jxo} for two-flavor $\chi$PT with a matching between two-flavor and  three-flavor  $\chi$PT, and genuine three-flavor $\chi$PT in Ref.~\cite{Lu:2020rhp}. These give values of $-E + 1.92(4)$ and $-E + 2.05(3)$, respectively, superseding our LO result. 

The contributions for large ALP masses have  been calculated for small isospin breaking in Refs.~\cite{DallaValleGarcia:2023xhh, Aghaie:2024jkj} for pure derivative couplings  ($E = N  = 0$), and for pure gauge bosons couplings ($c_i  = 0$) in Refs.~\cite{Ertas:2020xcc, Ovchynnikov:2025gpx}. When restricting our result in Appendix B.1 to the small isospin limit, we only agree with Ref.~\cite{Ovchynnikov:2025gpx}. In Ref.~\cite{Aghaie:2024jkj} there seems to be an apparent confusion of mixing angles of $\eta_0, \eta_8$ and $\eta, \eta^\prime$, while there is a discrepancy already on the level of mixing angles with Ref.~\cite{Ertas:2020xcc}, and their final result seems off by a   large ${\cal O}(1)$ factor for sizable values of $m_a$. With respect to Ref.~\cite{DallaValleGarcia:2023xhh} we find a different sign for the $\eta-a$ mixing angle. 

\item{\bf Light Pseudoscalar}

Our results directly apply to a light pseudoscalar $A$ with diagonal fermion couplings of the form ${\cal L} \supset - A/\Lambda \overline{f}_i i y_i \gamma_5 f_i$, giving $C_\gamma^{\rm eff} (A) = 0$ and  for $m_A \ll m_\pi$ 
\begin{align}
\tilde{C}_\gamma^\text{eff}(0)&=
\sum_{i=\mathrm{all}} N_C^i Q_i^2\, \frac{y_i}{m_i}- \frac{1}{2}    \left(\frac{5}{3} +  \delta \right) \sum_{i=\mathrm{quarks}}  \frac{y_i}{m_i} \, , 
\end{align}
where we also took the $m_\eta \to \infty$ limit for simplicity. Compared to the naive loop contribution from the first term, all quark contribution receives an ${\cal O}(1)$ correction due to the effective gluon coupling $\tilde{C}_G$, which is formally a 2-loop effect but of the same size since $\alpha_s$ is non-perturbative at the relevant scale $\mu = m_A$. This is in  contrast to the statements in Appendix B of Ref.~\cite{Dolan:2014ska}, which only considers  the first contribution.

\item{\bf CPon}

Spinless particle are expected to arise from solutions to the strong CP problem based on spontaneous CP violation~\cite{Nelson:1983zb,Barr:1984qx}.
The CKM phase arises as a VEV of a complex scalar field, a component of which --- the CPon --- can remain light
and provide a candidate of Dark Matter~\cite{Feruglio:2024dnc}. In the simplest version of the theory,
the couplings of the CPon $\phi$ are specified by~\cite{Feruglio:2023uof,Feruglio:2024ytl}:
\begin{align}
c_\gamma=c_G=\tilde c_\gamma=\tilde c_G=\xi=0\,,~~~~~~~~~~~~\sum \hat x_i=\sum \hat y_i=0\,,
\end{align}
where the last relations hold for sums restricted to each charge sector: charged leptons, up-type quarks and down-type quarks. This implies:
\begin{align}
\kappa & =\frac{2}{27}(\hat x_u+\hat x_d+\hat x_s)\,, & \tilde C_G & =0\,.
\end{align}
As a result, $\tilde{C}_\gamma^\text{eff}(m_\phi)$ is suppressed by powers $m_\phi^2/m_{\pi,\eta}^2$,
or $m_\phi^2/m_{e,\mu}^2$ and we get:
\begin{align}
C_\gamma^\text{eff}(m_\phi)&=
\frac{1}{81} (79 \hat x_u+ 25 \hat x_d+ 25\hat x_s) -\frac{1}{12}\left(\frac{x_{uu}+x_{dd}}{m_u+m_d}+\frac{x_{uu}+x_{ss}}{m_u+m_s}\right) \nn \\[0.3 cm]
&-\frac{7}{180}\sum_{i=e,\mu}\hat{x}_i
\frac{m_\phi^2}{m_i^2}
+\frac{17}{2430}(\hat x_u+\hat x_d+\hat x_s)\left(\frac{m_\phi^2}{m_\pi^2}+\frac{m_\phi^2}{m_K^2}\right)\nn\\
&-\frac{1}{90}\left(\frac{x_{uu}+x_{dd}}{m_u+m_d}\frac{m_\phi^2}{m_\pi^2}+\frac{x_{uu}+x_{ss}}{m_u+m_s}\frac{m_\phi^2}{m_K^2}\right)+ \hdots \\[0.3 cm]
\tilde{C}_\gamma^\text{eff}(m_\phi)&=\frac{1}{12}\sum_{i=e,\mu}\hat{y}_i\frac{m_\phi^2}{m_i^2}
 + \frac{  \hat y_{u} - \hat y_{d}}{2}\frac{m_\phi^2}{m_\pi^2 - m_\phi^2} \nn\\
& + \frac{m_\phi^2}{24m_\eta^2} \left[\left( 13 \hat y_{u} + 13 \hat y_{d} - 6 \hat y_{s}  \right)-  \delta\, \frac{m_\pi^2}{(m_\pi^2 - m_\phi^2)} \left( 4 \hat y_{d} + 6 \hat y_{s} - 22 \hat y_{u}  \right) \right]+ \hdots 
\end{align}
up to higher powers of $m_\phi^2$.
This extends the results of Ref.~\cite{Feruglio:2024dnc}, where only the limit $m_\phi=0$ was computed.
\end{itemize}

\section{Conclusions}
\label{sec:conclusions}
We have evaluated the effective couplings of a light spinless gauge-singlet particle $\phi$ to on-shell photons.
We considered the most general case of CP-violating microscopic interactions, allowing for both
scalar and pseudoscalar elementary couplings. By conventionally assigning $\phi$ to the CP-even sector,
we have formally distinguished between CP-invariant and CP-violating interactions. We focused on a light, weakly coupled particle with $m_\phi \ll {\rm GeV}$, whose interactions with matter,
the Higgs boson, and gauge fields are suppressed by inverse powers of a heavy scale $\Lambda$.

In specific models, such a particle may represent a viable dark matter candidate.
Stability on cosmological scales requires its mass to lie below the threshold for decay into an electron--positron pair, unless the couplings to electrons is extremely suppressed.
Moreover, the decay rate into two photons must be sufficiently small to evade stringent constraints,
such as those arising from stellar cooling and from indirect X-ray and gamma-ray line searches.
In general, a comprehensive assessment of the viable parameter space cannot be achieved without precise knowledge
of the dependence of the relevant physical observables on all particle couplings.
For instance, in specific scenarios, the decay rate into two photons can be entirely determined by the microscopic couplings of the decaying particle to quarks, leptons, and gluons.

Starting from the most general theory that includes a spinless gauge-singlet particle at the electroweak scale,
we derived the corresponding low-energy effective theory below the GeV scale, where light quarks
and gluons are replaced by a set of pseudoscalar mesons.
Using this effective theory, we systematically expanded the effective couplings of $\phi$ to on-shell photons
in powers of small parameters. Working at the one-loop level, we retained terms at first order in $1/\Lambda$.
We accounted for both isospin-breaking effects and $\eta$--$\eta'$ mixing, providing explicit expressions
for the couplings up to first order in $m_\phi^2/m_\eta^2$ and $m_\pi^2/m_\eta^2$.

The coupling of light scalars and pseudoscalars to photons at low energies has been extensively studied in the literature.
Nevertheless, we have extended existing analyses in several important aspects.
The decay of a hypothetical light Higgs particle --- with a mass below 1~GeV --- into two photons was investigated
in Ref.~\cite{Leutwyler:1989tn}, where the gluonic contribution was first evaluated using the QCD trace anomaly
and current algebra techniques.
Building on a similar theoretical framework, we have extended the analysis of Ref.~\cite{Leutwyler:1989tn}
to fully general CP-invariant couplings, including the possibility of mixing between $\phi$ and the Higgs boson as
a particular case.
Our treatment extends the recent analysis of Ref.~\cite{Flambaum:2024zyt} --- which also focuses on the CP-invariant sector ---
by fully incorporating the contribution arising from the $\phi$--gluon interaction.

In the CP-violating sector, we have comprehensively accounted for the effects of light pseudoscalar mesons,
including isospin-breaking effects and $\eta$--$\eta'$ mixing.
A crucial consistency check of our computation is provided by the independence of the final result under
the most general change of basis --- with three active light flavors, $u$, $d$, and $s$ --- designed to eliminate
the CP-violating $\phi$--gluon interaction.
In light of a discrepancy between our results and the analogous analysis of Ref.~\cite{Bai:2025fvl},
we explicitly demonstrate this independence using two distinct expansions, which are presented in the Appendix.
Our general results can be specialized to any physically motivated scenario involving a light spinless particle,
and we discuss several well-known examples in the final part of this work, pointing out various discrepancies with the existing literature.

\section*{Ackowledgments}
We thank Luca Di Luzio for useful dicussions.
G.L.~gratefully acknowledges financial support from the Swiss National Science Foundation (Project No.\ TMCG-2\_213690). F.F. work was supported by INFN. The work of R.Z.  is partially supported by project B3a of the DFG-funded Collaborative Research Center TRR257, ``Particle Physics Phenomenology after the Higgs Discovery" and has received support from the European Union's Horizon 2020 research and innovation programme under the Marie Sk\l{}odowska-Curie grant agreement No 860881-HIDDeN.

%%%%%%%%%%%%%%%%%%%%%%%%%%%%%%%%%%%%%%%%%%%%%%%%%%%%%%%%%
\begin{appendices}
%%%%%%%%%%%%%%%%%%%%%%%%%%%%%%%%%%%%%%%%%%%%%%%%%%%%%%%%%
\section{Operators involving $\phi$ and the Higgs boson}
\label{sec:basis}
%%%%%%%%%%%%%%%%%%%%%%%%%%%%%%%%%%%%%%%%%%%%%%%%%%%%%%%%%
Here we show that it is possible to choose a field basis where all derivative operators involving $\phi$ and the Higgs boson
are absent.
Operators containing derivatives have at least dimension five:
\begin{align}
O_1=&~\Box \phi\,|H|^2\,,&
O_2=&~ \phi\, \Box |H|^2\nn\\
O_3=&~\phi H^\dagger D^2 H\,,&
O_4=&~\phi D^2 H^\dagger\,H\nn\\
O_5=&~\phi D H^\dagger D H\,,&
O_6=&~\partial \phi D H^\dagger\,H\nn\\
O_7=&~\partial \phi H^\dagger\,D H\,.&
\end{align}
Via integration by parts, they are related by the equations
\begin{align}
O_1=&~O_2\,,&
O_2=&~O_3+O_4+2 O_5\nn\\
-O_6=&~O_3+O_5\,,&
-O_7=&~O_4+O_5\,,\nn
\end{align}
that can be used to eliminate the operators $O_2$, $O_5$, $O_6$ and $O_7$.
The operators $O_{3,4}$ can be removed by making use of the equations of motion of the Higgs doublet. In terms of the two combinations $O_\pm=O_3\pm O_4$ ($O_+$ is hermitian, while $O_-$ is antihermitian) we obtain:
\begin{align}
O_+=&~
2\phi(\lambda v^2 |H|^2-\lambda |H|^4)
-\phi (\bar q_L Y_d H d_R+\bar l_L Y_e H e_R+\bar q_L Y_u \tilde H u_R+h.c.)
\nn\\
O_-=&~
\phi (\bar q_L Y_d H d_R+\bar l_L Y_e H e_R-\bar q_L Y_u \tilde H u_R-h.c.)\,.\nn
\nn
\end{align}
Our effective Lagrangian already includes the operators (we use ${\cal Y}$ to distinguish these couplings from the Yukawas $Y$
of the SM):
\begin{align}
-\phi (\bar q_L {\cal Y}_d H d_R+\bar l_L  {\cal Y}_e H e_R+\bar q_L  {\cal Y}_u \tilde H u_R+h.c.)\,.\nn
\end{align}
The combination $c_+ O_++i c_- O_-$ produces a shift of the couplings ${\cal Y}$:
\begin{align}
{\cal Y}_{d,e}\to{\cal Y}_{d,e}+(c_+-ic_-) Y_{d,e}\,,~~~~~~~~{\cal Y}_u\to{\cal Y}_u+(c_++ic_-) Y_u\,.\nn
\end{align}
We conclude that $O_{3,4}$ can be eliminated in favour of the nonderivative operators
\be
\phi |H|^2\,,~~~\phi |H|^4\,.\nn
\ee
We are left with a single derivative operator, that can be chosen as $O_1$. To show that also $O_1$ is redundant,
consider the energy density
\begin{align}
V=&~V_H+V_\phi+V_{\phi H}\,.\nn
\end{align}
We first discuss $V_\phi$, the limit of $V$ when the Higgs doublet is set to zero. 
We assume the dependence of $V_\phi$ on $\phi$  is only through the combination
$(\phi/\Lambda)$:
\begin{align}
V_\phi=&~\mu^2\Lambda^2 f\left(\frac{\phi}{\Lambda}\right)=\mu^2 \Lambda^2\sum_{n=0}^\infty\frac{1}{n!}f^{(n)}(0)\frac{\phi^n}{\Lambda^n}\,.\nn
\end{align}
The function $f(x)$ is dimensionless and the overall scale $\mu^2\Lambda^2$
has been choosen to deliver a mass term for $\phi$ proportional to $\mu^2$, a free parameter. When $\mu^2=0$, the
potential is flat, reproducing the case of a Goldstone boson. When $\mu^2=\Lambda_{QCD}^4/\Lambda^2$,
an axion-like mass is reproduced. We have in mind an application where $\mu$ is below the electron-positron
energy threshold, so that $\phi$ can only decay into two photons or two neutrinos.
Neglecting all the $1/\Lambda$ terms, the leading equations of motion of $\phi$ read
\begin{align}
\Box\phi=~-f'(0)\mu^2 \Lambda-f''(0)\mu^2 \phi+...\nn
\end{align}
Replacing this in $O_1$ we obtain
\begin{align}
O_1=-f'(0)\mu^2 \Lambda|H|^2-f''(0)\mu^2 \phi |H|^2+...\nn
\end{align}
showing that $O_1$ can be absorbed by redefining $V_H+V_{\phi H}$. We write
\begin{align}
V_H=&~-\lambda v^2 |H|^2+\lambda |H|^4\nn\\
V_{\phi H}=&~\beta_1\frac{\phi}{\Lambda}v^2 |H|^2+ \beta_2 \frac{\phi}{\Lambda} |H|^4+{\cal O}(\phi^2/\Lambda^2)\,,\nn
\end{align}
understanding that 
the parameters $v^2$ and $\beta_1$ include the effect of $O_1$, respectively. Here again
we have assumed that the interaction of $\phi$ with the SM fields occurs via the combination
$(\phi/\Lambda)^k$ and we have kept only the first order $k=1$. In $V_H+V_{\phi H}$ we retained
operators up to dimension 5. 
It is not restrictive to choose a field basis where $\langle \phi\rangle=0$ (otherwise we conveniently shift $\phi$).
This imposes a relation on the minimum conditions. Setting $|H|=\rho/\sqrt{2}$,
\begin{align}
\frac{\partial V}{\partial \rho}=&~-\lambda v^2 \rho+\lambda \rho^3+ \beta_1\frac{\phi}{\Lambda}v^2 \rho+ \beta_2 \frac{\phi}{\Lambda} \rho^3\nn\\
=&~-\lambda\rho(v^2 -\rho^2)=0\nn\\
\frac{\partial V}{\partial \phi}=&~\mu^2 \Lambda\sum_{n=0}^\infty\frac{1}{n!}f^{(n+1)}(0)\frac{\phi^{n}}{\Lambda^{n}}
+\beta_1\frac{1}{2\Lambda}v^2 \rho^2+ \beta_2 \frac{1}{4\Lambda} \rho^4\nn\\
=&~ f'(0)\mu^2\Lambda+\beta_1\frac{1}{2\Lambda}v^2 \rho^2+ \beta_2 \frac{1}{4\Lambda} \rho^4=0\,.\nn
\end{align}
These are solved by
\begin{align}
\langle H\rangle=&~\frac{v}{\sqrt{2}}\nn\\
f'(0)\mu^2=&~-(2\beta_1+\beta_2)\frac{v^4}{4\Lambda^2}\,.\nn
\end{align}
Evaluating the second derivatives at the minimum we find
the mass matrix
\begin{align}
\left(
\begin{array}{cc}
2 \lambda v^2&(\beta_1+\beta_2)\frac{v^3}{\Lambda}\\
(\beta_1+\beta_2)\frac{v^3}{\Lambda}&f''(0) \mu^2
\end{array}
\right)\,.\nn
\end{align}
To first order in $v/\Lambda$ and in the limit $f''(0) \mu^2\ll \lambda v^2$, the mass eigenstates are:
\begin{align}
\left(
\begin{array}{c}
\hat H\\\hat \phi
\end{array}
\right)=\left(
\begin{array}{cc}
1&\xi\dd\frac{v}{\Lambda}\\
-\xi\dd\frac{v}{\Lambda}&1
\end{array}
\right)
\left(
\begin{array}{c}
H\\\phi
\end{array}
\right)\,,\nn
\end{align}
where $\xi=\frac{(\beta_1+\beta_2)}{2\lambda}$ is the mixing parameter used in the text.

In summary, all derivative $\phi-H$ interactions can be eliminated in favour of non-derivatives ones. At order $1/\Lambda$,
these are $\phi |H|^2$ and $\phi |H|^4$. Expanding the energy density around the minimum, the main effect
is represented by a mixing between $\phi$ and $H$, of order $v/\Lambda$, contributing to the
$\phi\to\gamma\gamma$ amplitude.

\section{$\phi$-meson mixing } 
\label{sec:mixing}
In the basis $P=(\phi,\pi^0,\eta_8,\eta_0)$, the quadratic part of the $\phi$-meson Lagrangian reads 
\be\nn
L_2=\frac{1}{2} (\partial_\mu P)^T K_P (\partial^\mu P) -\frac{1}{2}P^T M_P^2 P\,,
\ee
where, setting $\epsilon=f_\pi/\Lambda$, the matrices $K_P$ and $M_P^2$ are given by
\begin{align}
K_P & = \left(
\begin{array}{cccc}
1&c_1\epsilon&c_2\epsilon&c_3\epsilon\\
c_1\epsilon&1&0&0\\
c_2\epsilon&0&1&0\\
c_3\epsilon&0&0&1
\end{array}
\right)\,, \\
M_P^2 & = \left(
\begin{array}{cccc}
m_\phi^2&-m_{12}^2 \epsilon&-m_{13}^2 \epsilon&-m_{14}^2 \epsilon\\
-m_{12}^2 \epsilon&B_0(m_u+m_d)&\frac{\dd B_0}{\sqrt{3}}(m_u-m_d)&B_0\sqrt{\frac23}(m_u-m_d)\\
-m_{13}^2 \epsilon&\frac{\dd B_0}{\sqrt{3}}(m_u-m_d)&m_{\eta_8}^2&\delta m^2\\
-m_{14}^2 \epsilon&B_0\sqrt{\frac23}(m_u-m_d)&\delta m^2&m_{\eta_0}^2\\
\end{array}
\right)\, , 
\end{align}
with
\begin{align}
c_1 & = \frac12\left(\hat y_{u}-\hat y_{d}\right)-2\tilde C_G(Q_u-Q_d)\,,\nn\\
c_2 & = \frac{1}{\sqrt{3}}\left[\frac12\left(\hat y_{u}+\hat y_{d}-2\hat y_{s}\right)-2\tilde C_G(Q_u+Q_d-2Q_s)\right]\,,\nn\\
c_3 & = \sqrt{\frac23}\left[\frac12\left(\hat y_{u}+\hat y_{d}+\hat y_{s}\right)-2\tilde C_G(Q_u+Q_d+Q_s)\right]\,,\nn \\
\nn \\
m_{12}^2& = 4\tilde C_G B_0(m_u Q_u-m_d Q_d)\,,& m_{\eta_8}^2&=\frac{B_0}{3}(m_u+m_d+4m_s)\,, \nn\\
m_{13}^2& = \frac{4}{\sqrt{3}}\tilde C_G B_0(m_u Q_u+m_d Q_d-2m_s Q_s)\, , & m_{\eta_0}^2&=\frac{2B_0}{3}(m_u+m_d+m_s)+M_0^2\,,\nn\\
m_{14}^2 & =  \frac{4\sqrt{2}}{\sqrt{3}}\tilde C_G B_0(m_u Q_u+m_d Q_d+m_s Q_s)\,, & \delta m^2&=\frac{\sqrt{2}}{3}B_0(m_u+m_d-2m_s) \,,
\end{align}
where we recall that in this normalization $Q_u + Q_d + Q_s = {\rm tr} \, Q_A = 1/2$.

One could in principle perform an exact diagonalization of these matrices to identify the canonically normalized mass eigenstates $(\phi_{\rm phys},\pi^0_{\rm phys}, \eta_{\rm phys}, \eta_{\rm phys}^\prime)$. However, it is well known that $\chi PT$ at leading order is not adequate to describe $\eta - \eta^\prime$ mixing \cite{Georgi:1993jn, Gerard:2004gx}, and 
 corrections to mixing angles
and masses arise at NLO~\cite{Leutwyler:1997yr, Beisert:2001qb, Alves:2017avw}. Here we ignore these corrections, which justifies to  perform a diagonalization only at leading order in the small parameters $w \equiv m_u/m_s \sim 0.02$ and $\delta \equiv (m_d - m_u)/(m_d + m_u)\sim 0.3$ (and $\eps = f_\pi/\Lambda$). 

We begin by moving from the $(\eta_8,\eta_0)$ basis to the $(\eta,\eta')$ basis, through the rotation
\begin{align}
\left(
\begin{array}{c}
\eta_8\\ \eta_0
\end{array}
\right)=
\left(
\begin{array}{cc}
\cos\alpha&\sin\alpha\\
-\sin\alpha&\cos\alpha
\end{array}
\right)
\left(
\begin{array}{c}
\eta\\ \eta'
\end{array}
\right)\,.
\end{align}
This diagonalizes the 3-4 block of the mass matrix, with diagonal entries that to the order we are interested in (up to multiplicative corrections of order ${\cal O}(\delta^2 w^2)$) can be identified with the physical masses $m_\eta^2$ and $m_{\eta^\prime}^2$. These masses and the mixing angle $\alpha$ are related to the entries in the 3-4 block ($\delta m^2, m_{\eta_8}, m_{\eta_0}$) by the three equations 
\begin{align}
\label{rot}
\tan2\alpha & =\frac{2\delta m^2}{m_{\eta_0}^2-m_{\eta_8}^2}\,,
& \cos2\alpha & = \frac{m_{\eta_8}^2-m_{\eta_0}^2}{m_\eta^2-m_{\eta'}^2} \,,
&  m_\eta^2+m_{\eta'}^2 & =m_{\eta_8}^2+m_{\eta_0}^2 \,.
\end{align}
In turn these relations allow us to express $M_0^2$ and $m_s$ in terms of the mixing angle $\alpha$ and $m_\eta^2$, giving (we introduce the shorthand $c \equiv \cos \alpha, s \equiv \sin \alpha$)
\begin{align}
M_0^2 & = 2 B_0 \left(\frac{m_u}{1-\delta}  -  m_s \right)  \frac{\sqrt2 (c^2 -s^2) - c s }{3 c s} \, ,  \\
B_0 m_s & = m_\eta^2  \frac{3 c  }{2 (2 c + \sqrt 2 s)} +  B_0 m_u \frac{c  - 
  \sqrt2   s}{  (-1 + \delta) (2 c + \sqrt 2 s)} \, , 
\end{align}
along with the relation
\begin{align}
m_{\eta^\prime}^2 & =m_\eta^2 \frac{2 c   s  - \sqrt 2c^2 }{s (2 c + \sqrt 2s)} - B_0 m_u \frac{ 
 2 \sqrt 2  }{(-1 + \delta) s (2 c + \sqrt 2s)} \, .
\end{align}
Finally we can express $B_0 m_u$ through the mass of the neutral pion $m_\pi^2$. The leading order relation is given by 
\begin{align}
\label{B0}
B_0 m_u & = \frac{1-\delta}{2} m_\pi^2\,,
\end{align}
and receives multiplicative corrections of order $\delta^2 w$. Therefore the LO expressions is sufficient to express 
all heavy scales through physical masses as (isospin breaking effects only enter at higher order)
\begin{align}
M_0^2 & =  \left( m_\eta^2 - m_\pi^2 \right) \frac{\sqrt{2} (s^2-c^2) + cs}{s(2c + \sqrt 2 s)} \, ,  \\
B_0 m_s & = m_\eta^2 \frac{3 c}{4 c + 2 \sqrt2 s} + m_\pi^2 \frac{\sqrt 2 s - c }{4c + 2 \sqrt2 s}\, , \\
  m_{\eta^\prime}^2 & =m_\eta^2 \frac{c (2  s - \sqrt2 c) }{s (2 c + \sqrt 2s)} + m_\pi^2  \frac{ 
 1  }{ s ( \sqrt 2 c +  s)} \, .
\end{align}
As anticipated, the last relation does not allow for realistic $\eta$- and $\eta^\prime$ masses, which require NLO $\chi$PT. Here we follow Refs.~\cite{Aloni:2018vki, Cheng:2021kjg} and make the choice  $s  = -1/3, c =2\sqrt{2}/3$, which  gives the simple expressions
\begin{align}
\label{massfinal}
M_0^2 & =  3 \left( m_\eta^2 - m_\pi^2 \right)  \, ,  & 
B_0 m_s & = m_\eta^2 - \frac{m_\pi^2}{2}  \, , &
  m_{\eta^\prime}^2 & =4 m_\eta^2 - 3 m_\pi^2 \, .
\end{align}
Taking the expressions in Eqs.~(\ref{massfinal}, \ref{B0}) along with $s  = -1/3, c =2\sqrt{2}/3$, one can calculate the mixing angles $\theta_{P \phi}$ of $\phi$ with the meson mass eigenstate $P$, which is the first column of the orthogonal matrix $O$ that diagonalizes the mass matrix in the basis where kinetic terms are diagonal, i.e.
\begin{align}
\theta_{P \phi} & = O_{P1} \, , & O^T R^T M^2_P R O & = M^2_{\rm diag} \, , & R^T K_P R & = \mathbb{1}_{4 \times 4} \, .
\end{align}
Plugging these mixing angles into the 
Wess-Zumino-Witten (WZW) term~\cite{Wess:1971yu, Witten:1983tw}:
\begin{equation}
{\cal L}_{WZW} \supset -\frac{\alpha}{4\pi f_\pi}\left(\pi^0+\frac{1}{\sqrt{3}}\eta_8+2\sqrt{\frac23}\eta_0\right)F_{\mu\nu}\tilde F^{\mu\nu}\,,
\end{equation}
or equivalently in the $\eta - \eta^\prime$ basis for the chosen value of $\alpha$, 
\begin{equation}
-\frac{\alpha}{4\pi f_\pi}\left(\pi^0+\frac{4 \sqrt{2/3}}{3} \eta + \frac{7}{3 \sqrt 3}\eta^\prime\right)F_{\mu\nu}\tilde F^{\mu\nu}\,,
\end{equation}
one obtains a contribution to the effective $\phi$-couplings to photons ${\cal L} \supset -\delta\tilde C_\gamma \alpha/4\pi \phi/\Lambda F_{\mu\nu}\tilde F^{\mu\nu}$ given by
\begin{align}
\delta\tilde C_\gamma & = \frac{1}{\eps} \sum_{P = \pi^0, \eta, \eta^\prime} C_P \theta_{P \phi}  \,, 
\end{align}
with $C_\pi  =  1, C_\eta = 4/3 \sqrt{2/3}, C_{\eta^\prime} = 7/(3 \sqrt 3)$. This adds to the one controlled by
\begin{equation}
\tilde{C}_\gamma= \tilde c_\gamma +\sum_{i=\mathrm{all}} N_C^i Q_i^2\,\hat y_i  - 4  \tilde{C}_G \sum_{i=u,d,s}N_C^i Q_i^2 Q_{Aii} \,, 
\end{equation}
such that the total contribution $\tilde C_\gamma+\delta\tilde C_\gamma$ should not depend on the choice of the matrix $Q_A$, which provides a useful consistency check. 

We now give the mixing angles in two expansion schemes: first we follow Ref.~\cite{Bai:2025fvl} and expand only in the isospin breaking parameter $\delta$ (which is actually not particularly small), second we expand in $w$ to first order (i.e. neglecting corrections ${\cal O}(m_{u,d}^2/m_s^2 \sim m_\pi^4/m_\eta^4$) and keep all orders of $\delta$. 

\subsection{Mixing angles in the small isospin-breaking Expansion}
\label{sec:deltaexpansion}
Expanding to first order in $\eps$ and $\delta = (m_d - m_u)/(m_d+m_u)$, the relevant mixing angles are given by (assuming that $m_\phi$ is not too close to $m_\pi, m_\eta$ or $m_{\eta^\prime}$)

\begin{align}
\theta_{\pi\phi}/\epsilon & = 2  \tilde{C}_G  (Q_u - Q_d) + \frac{m_\phi^2}{m_\pi^2 - m_\phi^2} \frac{\hat y_{u} - \hat y_{d}}{2}
 -2 \delta \tilde{C}_G   \frac{m_\pi^2(m_\eta^2-m_\pi^2) (2 m_\eta^2-m_\pi^2-m_\phi^2) }{(m_\eta^2 - m_\phi^2) (m_{\eta'}^2 - m_\phi^2) (m_\pi^2-m_\phi^2)}  \nn \\ 
 & -  \delta \frac{m_\phi^2 m_\pi^2 \left(m_\phi^2 (\hat y_{u} + \hat y_{d}) + m_\eta^2 (- 3 \hat y_{u} -3\hat y_{d} + 2 \hat y_{s} ) + 2m_\pi^2 (\hat y_{u} + \hat y_{d} - \hat y_{s})\right)}{2(m_\eta^2 - m_\phi^2) (m_{\eta'}^2 - m_\phi^2) (m_\pi^2-m_\phi^2)} \, , \\
\theta_{\eta\phi}/\epsilon & = \frac{\sqrt{2/3} }{m_\eta^2 - m_\phi^2} \left[  m_\phi^2 \left( - 2  \tilde{C}_G (Q_u + Q_d -  Q_s) + \frac{\hat y_{u} + \hat y_{d} - \hat y_{s}}{2} \right) +  m_\pi^2  \tilde{C}_G  - 4 m_\eta^2 \tilde{C}_G Q_s\right]    \\
& +  \delta \frac{\sqrt{2/3} \, m_\phi^2 m_\pi^2 }{2(m_\eta^2 - m_\phi^2)(m_\pi^2 - m_\phi^2)} (\hat y_{u} - \hat y_{d})\, , \nn \\
\theta_{\eta^\prime\phi}/\epsilon & = \frac{\sqrt{1/3} }{m_{\eta^\prime}^2 - m_\phi^2} \left[   m_\phi^2 \left( - 2  \tilde{C}_G (Q_u + Q_d +  2 Q_s) +\frac{\hat y_{u} + \hat y_{d} + 2 \hat y_{s}}{2} \right) \right. \nn \\ & \left. + 2 m_\pi^2  \tilde{C}_G (Q_u + Q_d  - 2 Q_s)  + 8 m_\eta^2 \tilde{C}_G Q_s \right]  +  \delta \frac{\sqrt{1/3} \, m_\phi^2 m_\pi^2 }{2(m_{\eta^\prime}^2 - m_\phi^2)(m_\pi^2 - m_\phi^2)} (\hat y_{u} - \hat y_{d}) \, .
\end{align}
These results agree with Eq.~(3.15) in Ref.~\cite{Bai:2025fvl} only in the isopsin limit, i.e. for  $\delta \to 0$.  Calculating $\delta\tilde C_\gamma$ and summing with  $\tilde{C}_{\gamma G} \equiv - 4  \tilde{C}_G \sum_{i=u,d,s}N_C^i Q_i^2 Q_{Aii}$, this contribution does indeed not depend on $Q_A$, and reads
\begin{align}
\label{delta}
\delta \tilde{C}_\gamma  + \tilde{C}_{\gamma G} &=- \frac53 \tilde C_G+\frac89 \tilde C_G\frac{m_\pi^2-m_\phi^2}{m_\eta^2-m_\phi^2}+\frac79 \tilde C_G\frac{m_\pi^2-m_\phi^2}{m_{\eta'}^2-m_\phi^2}
\nn\\
& + \frac{4}{9} \left( \hat y_{u} + \hat y_{d} - \hat y_{s} \right)  \frac{ m_\phi^2}{m_\eta^2 - m_\phi^2}+ \frac{ \hat y_{u} - \hat y_{d}}{2}   \frac{ m_\phi^2}{m_\pi^2 - m_\phi^2} \nn \\ & + \frac{7}{18} \left( \hat y_{u} + \hat y_{d} +2 \hat y_{s} \right)  \frac{ m_\phi^2}{m_{\eta^\prime}^2 - m_\phi^2} \nn \\
&+\delta \tilde C_G\left[-\frac{m_\pi^2}{m_\pi^2-m_\phi^2}+\frac23\frac{m_\pi^2}{m_\eta^2-m_\phi^2}+\frac13\frac{m_\pi^2}{m_{\eta'}^2-m_\phi^2}
\right] \nn \\
 &  -\frac{\delta}{3}m_\pi^2m_\phi^2 \left[\frac{7 \hat y_{u}-\hat y_{d}-3\hat y_{s}}{(m_\eta^2-m_\phi^2)(m_{\eta'}^2-m_\phi^2)}-\frac{11 \hat y_{u}-2\hat y_{d}-3\hat y_{s}}{(m_\pi^2-m_\phi^2)(m_{\eta'}^2-m_\phi^2)}\right]  \,.  
\end{align}
This agrees with the result in Eq.~(3.18) of Ref.~\cite{Bai:2025fvl} for all terms except the ones proportional to $\tilde{C}_G$ in the isospin limit.

\subsection{Mixing angles in the large \boldmath${m_s}$ Expansion}
Expanding to first order in $\eps$ and $m_{u,d}/m_s \sim m_\pi^2/m_\eta^2$ (and  to first order in  $m_\phi^2/m_\eta^2$), the relevant mixing angles are given by (provided that $m_\phi$ is not too close to $m_\pi$)
\begin{align}
 \theta_{\pi \phi}/\eps & =\frac{m_\phi^2}{m_\pi^2 - m_\phi^2} \left( \frac{\hat y_{u} - \hat y_{d}}{2} + 2  \tilde{C}_G (Q_d - Q_u) \right) + \frac{ m_\pi^2}{m_\pi^2 - m_\phi^2}  \tilde{C}_G \left( 2Q_u - 2Q_d  - \delta \right)  \nn \\
 & +  \frac{\delta m_\phi^2 m_\pi^2}{8 m_\eta^2 (m_\pi^2 - m_\phi^2)} \left( 3 \hat y_{u} + 3 \hat y_{d} - 2 \hat y_{s} - 6  \tilde{C}_G  \right) +  \frac{3 \delta (1- \delta^2) m_\pi^4}{4 m_\eta^2 (m_\pi^2 - m_\phi^2)}   \tilde{C}_G  \, , \\
\sqrt{\frac{3}{2}}  \theta_{\eta \phi}/\eps & = -4  \tilde{C}_G Q_s +   \frac{m_\phi^2}{2 m_\eta^2}  \left( \hat y_{u} +\hat y_{d}  - \hat y_{s}-  2\tilde{C}_G \right)+   \frac{m_\pi^2}{m_\eta^2}    \tilde{C}_G \nonumber \\
& +  \frac{\delta m_\phi^2 m_\pi^2 }{2 m_\eta^2 (m_\pi^2 - m_\phi^2)} \left( \hat y_{u} - \hat y_{d} \right) -    \frac{\delta^2 m_\pi^4 }{m_\eta^2 (m_\pi^2 - m_\phi^2)} \tilde{C}_G \, , \\
\sqrt{3}  \theta_{\eta^\prime \phi}/\eps & = 2 \tilde{C}_G Q_s +   \frac{m_\phi^2}{8 m_\eta^2}  \left( \hat y_{u} + \hat y_{d}+ 2 \hat y_{s} -  2 \tilde{C}_G \right)+   \frac{m_\pi^2}{4 m_\eta^2}    \tilde{C}_G \nonumber \\
& +  \frac{\delta m_\phi^2 m_\pi^2 }{8 m_\eta^2 (m_\pi^2 - m_\phi^2)} \left( \hat y_{u} - \hat y_{d} \right) -    \frac{ \delta^2 m_\pi^4 }{4 m_\eta^2 (m_\pi^2 - m_\phi^2)} \tilde{C}_G \, .
\end{align}
Calculating $\delta\tilde C_\gamma$ and summing with  $\tilde{C}_{\gamma G} = - 4  \tilde{C}_G \sum_{i=u,d,s}N_C^i Q_i^2 Q_{Aii}$, this contribution does indeed not depend on $Q_A$, and reads
\begin{align}
\label{RZ}
\delta\tilde C_\gamma + \tilde{C}_{\gamma G} & = - \frac{5}{3} \tilde{C}_G + \frac{  \hat y_{u} - \hat y_{d}}{2}\frac{m_\phi^2}{m_\pi^2 - m_\phi^2} -  \delta\,   \tilde{C}_G  \frac{m_\pi^2}{m_\pi^2 - m_\phi^2}  \nn \\
& + \frac{m_\phi^2}{24m_\eta^2} \left( 13 \hat y_{u} + 13 \hat y_{d} - 6 \hat y_{s} - 26 \tilde{C}_G \right) + \frac{13 m_\pi^2}{12m_\eta^2}  \tilde{C}_G \nn \\
& -  \delta\, \frac{ m_\phi^2 m_\pi^2}{24m_\eta^2 (m_\pi^2 - m_\phi^2)} \left( 4 \hat y_{d} + 6 \hat y_{s} - 22 \hat y_{u} + 18  \tilde{C}_G \right)  \nn \\ 
& + \delta\, ( 9 - 13 \delta - 9 \delta^2  ) \frac{ m_\pi^4}{12m_\eta^2 (m_\pi^2 - m_\phi^2 )}   \tilde{C}_G \, .
\end{align}
Further expanding to first order in $\delta$ matches the result in Eq.~(\ref{delta}), when expanded to first order in $1/m_\eta^2$.  In the $m_{\eta} \to \infty$ limit one obtains
\begin{align}
\delta \tilde{C}_\gamma  + \tilde{C}_{\gamma G} &=- \frac53 \tilde C_G + \frac{ \hat y_{u} - \hat y_{d}  }{2} \frac{ m_\phi^2}{m_\pi^2 - m_\phi^2}  - \delta \tilde C_G \frac{m_\pi^2}{m_\pi^2-m_\phi^2} \, , \end{align}
in agreement with Eq.~(92) of Ref.~\cite{Bauer:2020jbp}.
 In the main text we have adopted the contribution in Eq.~(\ref{RZ}) to derive the total result in Eq.~\eqref{tildeCgamma}.
\end{appendices}

\begin{small}
\bibliographystyle{utphys}
\bibliography{Bibliography}

@article{Peccei:1977hh,
    author = "Peccei, R. D. and Quinn, Helen R.",
    title = "{CP Conservation in the Presence of Instantons}",
    reportNumber = "ITP-568-STANFORD",
    doi = "10.1103/PhysRevLett.38.1440",
    journal = "Phys. Rev. Lett.",
    volume = "38",
    pages = "1440--1443",
    year = "1977"
}

@article{Peccei:1977ur,
    author = "Peccei, R. D. and Quinn, Helen R.",
    title = "{Constraints Imposed by CP Conservation in the Presence of Instantons}",
    reportNumber = "ITP-572-STANFORD",
    doi = "10.1103/PhysRevD.16.1791",
    journal = "Phys. Rev. D",
    volume = "16",
    pages = "1791--1797",
    year = "1977"
}

@article{Wilczek:1977pj,
    author = "Wilczek, Frank",
    title = "{Problem of Strong  $P$  and  $T$  Invariance in the Presence of Instantons}",
    reportNumber = "Print-77-0939 (COLUMBIA)",
    doi = "10.1103/PhysRevLett.40.279",
    journal = "Phys. Rev. Lett.",
    volume = "40",
    pages = "279--282",
    year = "1978"
}

@article{Weinberg:1977ma,
    author = "Weinberg, Steven",
    title = "{A New Light Boson?}",
    reportNumber = "HUTP-77/A074",
    doi = "10.1103/PhysRevLett.40.223",
    journal = "Phys. Rev. Lett.",
    volume = "40",
    pages = "223--226",
    year = "1978"
}

@article{Graham:2015cka,
    author = "Graham, Peter W. and Kaplan, David E. and Rajendran, Surjeet",
    title = "{Cosmological Relaxation of the Electroweak Scale}",
    eprint = "1504.07551",
    archivePrefix = "arXiv",
    primaryClass = "hep-ph",
    doi = "10.1103/PhysRevLett.115.221801",
    journal = "Phys. Rev. Lett.",
    volume = "115",
    number = "22",
    pages = "221801",
    year = "2015"
}

@article{Wilczek:1982rv,
    author = "Wilczek, Frank",
    title = "{Axions and Family Symmetry Breaking}",
    doi = "10.1103/PhysRevLett.49.1549",
    journal = "Phys. Rev. Lett.",
    volume = "49",
    pages = "1549--1552",
    year = "1982"
}

@article{Calibbi:2016hwq,
    author = "Calibbi, Lorenzo and Goertz, Florian and Redigolo, Diego and Ziegler, Robert and Zupan, Jure",
    title = "{Minimal axion model from flavor}",
    eprint = "1612.08040",
    archivePrefix = "arXiv",
    primaryClass = "hep-ph",
    reportNumber = "TTP16-058, CERN-TH-2016-261",
    doi = "10.1103/PhysRevD.95.095009",
    journal = "Phys. Rev. D",
    volume = "95",
    number = "9",
    pages = "095009",
    year = "2017"
}

@article{Ema:2016ops,
    author = "Ema, Yohei and Hamaguchi, Koichi and Moroi, Takeo and Nakayama, Kazunori",
    title = "{Flaxion: a minimal extension to solve puzzles in the standard model}",
    eprint = "1612.05492",
    archivePrefix = "arXiv",
    primaryClass = "hep-ph",
    reportNumber = "UT-16-36, IPMU16-0189",
    doi = "10.1007/JHEP01(2017)096",
    journal = "JHEP",
    volume = "01",
    pages = "096",
    year = "2017"
}

@article{Chikashige:1980ui,
    author = "Chikashige, Y. and Mohapatra, Rabindra N. and Peccei, R. D.",
    title = "{Are There Real Goldstone Bosons Associated with Broken Lepton Number?}",
    reportNumber = "MPI-PAE-PTH-36-80",
    doi = "10.1016/0370-2693(81)90011-3",
    journal = "Phys. Lett. B",
    volume = "98",
    pages = "265--268",
    year = "1981"
}

@article{Schechter:1981cv,
    author = "Schechter, J. and Valle, J. W. F.",
    title = "{Neutrino Decay and Spontaneous Violation of Lepton Number}",
    reportNumber = "SU-4217-203, COO-3533-203",
    doi = "10.1103/PhysRevD.25.774",
    journal = "Phys. Rev. D",
    volume = "25",
    pages = "774",
    year = "1982"
}

@article{Salam:1969bwb,
    author = "Salam, Abdus and Strathdee, J. A.",
    title = "{Nonlinear realizations. 2. Conformal symmetry}",
    reportNumber = "IC-68-107",
    doi = "10.1103/PhysRev.184.1760",
    journal = "Phys. Rev.",
    volume = "184",
    pages = "1760--1768",
    year = "1969"
}

@article{Ellis:1970yd,
    author = "Ellis, John R.",
    title = "{Aspects of conformal symmetry and chirality}",
    doi = "10.1016/0550-3213(70)90422-0",
    journal = "Nucl. Phys. B",
    volume = "22",
    pages = "478--492",
    year = "1970",
    note = "[Erratum: Nucl.Phys.B 25, 639--639 (1971)]"
}

@article{Goldberger:2007zk,
    author = "Goldberger, Walter D. and Grinstein, Benjamin and Skiba, Witold",
    title = "{Distinguishing the Higgs boson from the dilaton at the Large Hadron Collider}",
    eprint = "0708.1463",
    archivePrefix = "arXiv",
    primaryClass = "hep-ph",
    reportNumber = "UCSD-PTH-07-09",
    doi = "10.1103/PhysRevLett.100.111802",
    journal = "Phys. Rev. Lett.",
    volume = "100",
    pages = "111802",
    year = "2008"
}

@inproceedings{Cicoli:2013ana,
    author = "Cicoli, Michele",
    title = "{Axion-like Particles from String Compactifications}",
    booktitle = "{9th Patras Workshop on Axions, WIMPs and WISPs}",
    eprint = "1309.6988",
    archivePrefix = "arXiv",
    primaryClass = "hep-th",
    doi = "10.3204/DESY-PROC-2013-04/cicoli_michele",
    pages = "235--242",
    year = "2013"
}

@article{Randall:1999ee,
    author = "Randall, Lisa and Sundrum, Raman",
    title = "{A Large mass hierarchy from a small extra dimension}",
    eprint = "hep-ph/9905221",
    archivePrefix = "arXiv",
    reportNumber = "MIT-CTP-2860, PUPT-1860, BUHEP-99-9",
    doi = "10.1103/PhysRevLett.83.3370",
    journal = "Phys. Rev. Lett.",
    volume = "83",
    pages = "3370--3373",
    year = "1999"
}

@article{Goldberger:1999uk,
    author = "Goldberger, Walter D. and Wise, Mark B.",
    title = "{Modulus stabilization with bulk fields}",
    eprint = "hep-ph/9907447",
    archivePrefix = "arXiv",
    reportNumber = "CALT-68-2232",
    doi = "10.1103/PhysRevLett.83.4922",
    journal = "Phys. Rev. Lett.",
    volume = "83",
    pages = "4922--4925",
    year = "1999"
}

@article{Goldberger:1999un,
    author = "Goldberger, Walter D. and Wise, Mark B.",
    title = "{Phenomenology of a stabilized modulus}",
    eprint = "hep-ph/9911457",
    archivePrefix = "arXiv",
    reportNumber = "CALT-68-2250",
    doi = "10.1016/S0370-2693(00)00099-X",
    journal = "Phys. Lett. B",
    volume = "475",
    pages = "275--279",
    year = "2000"
}

@inproceedings{Giannotti:2015dwa,
    author = "Giannotti, Maurizio",
    title = "{ALP hints from cooling anomalies}",
    booktitle = "{11th Patras Workshop on Axions, WIMPs and WISPs}",
    eprint = "1508.07576",
    archivePrefix = "arXiv",
    primaryClass = "astro-ph.HE",
    doi = "10.3204/DESY-PROC-2015-02/giannotti_maurizio",
    pages = "26--30",
    year = "2015"
}

@article{Giannotti:2017hny,
    author = "Giannotti, Maurizio and Irastorza, Igor G. and Redondo, Javier and Ringwald, Andreas and Saikawa, Ken'ichi",
    title = "{Stellar Recipes for Axion Hunters}",
    eprint = "1708.02111",
    archivePrefix = "arXiv",
    primaryClass = "hep-ph",
    reportNumber = "DESY-17-116",
    doi = "10.1088/1475-7516/2017/10/010",
    journal = "JCAP",
    volume = "10",
    pages = "010",
    year = "2017"
}

@article{Saikawa:2019lng,
    author = "Saikawa, Ken'ichi and Yanagida, Tsutomu T.",
    title = "{Stellar cooling anomalies and variant axion models}",
    eprint = "1907.07662",
    archivePrefix = "arXiv",
    primaryClass = "hep-ph",
    reportNumber = "MPP-2019-130",
    doi = "10.1088/1475-7516/2020/03/007",
    journal = "JCAP",
    volume = "03",
    pages = "007",
    year = "2020"
}

@article{Badziak:2021apn,
    author = "Badziak, Marcin and Grilli di Cortona, Giovanni and Tabet, Mustafa and Ziegler, Robert",
    title = "{Flavor-violating Higgs decays and stellar cooling anomalies in axion models}",
    eprint = "2107.09708",
    archivePrefix = "arXiv",
    primaryClass = "hep-ph",
    reportNumber = "TTP21-025, P3H-21-051",
    doi = "10.1007/JHEP10(2021)181",
    journal = "JHEP",
    volume = "10",
    pages = "181",
    year = "2021"
}

@article{DiLuzio:2021ysg,
    author = "Di Luzio, Luca and Fedele, Marco and Giannotti, Maurizio and Mescia, Federico and Nardi, Enrico",
    title = "{Stellar evolution confronts axion models}",
    eprint = "2109.10368",
    archivePrefix = "arXiv",
    primaryClass = "hep-ph",
    reportNumber = "DESY-21-141, TTP21-030, P3H-21-062",
    doi = "10.1088/1475-7516/2022/02/035",
    journal = "JCAP",
    volume = "02",
    number = "02",
    pages = "035",
    year = "2022"
}

@article{Abbott:1982af,
    author = "Abbott, L. F. and Sikivie, P.",
    editor = "Srednicki, M. A.",
    title = "{A Cosmological Bound on the Invisible Axion}",
    reportNumber = "PRINT-82-0695 (BRANDEIS)",
    doi = "10.1016/0370-2693(83)90638-X",
    journal = "Phys. Lett. B",
    volume = "120",
    pages = "133--136",
    year = "1983"
}

@article{Dine:1982ah,
    author = "Dine, Michael and Fischler, Willy",
    editor = "Srednicki, M. A.",
    title = "{The Not So Harmless Axion}",
    reportNumber = "UPR-0201T",
    doi = "10.1016/0370-2693(83)90639-1",
    journal = "Phys. Lett. B",
    volume = "120",
    pages = "137--141",
    year = "1983"
}

@article{Preskill:1982cy,
    author = "Preskill, John and Wise, Mark B. and Wilczek, Frank",
    editor = "Srednicki, M. A.",
    title = "{Cosmology of the Invisible Axion}",
    reportNumber = "HUTP-82-A048, NSF-ITP-82-103",
    doi = "10.1016/0370-2693(83)90637-8",
    journal = "Phys. Lett. B",
    volume = "120",
    pages = "127--132",
    year = "1983"
}

@article{Hall:2009bx,
    author = "Hall, Lawrence J. and Jedamzik, Karsten and March-Russell, John and West, Stephen M.",
    title = "{Freeze-In Production of FIMP Dark Matter}",
    eprint = "0911.1120",
    archivePrefix = "arXiv",
    primaryClass = "hep-ph",
    reportNumber = "OUTP-09-18-P, UCB-PTH-09-32",
    doi = "10.1007/JHEP03(2010)080",
    journal = "JHEP",
    volume = "03",
    pages = "080",
    year = "2010"
}

@article{DiLuzio:2020oah,
    author = {Di Luzio, Luca and Gr{\"o}ber, Ramona and Paradisi, Paride},
    title = "{Hunting for $CP$-violating axionlike particle interactions}",
    eprint = "2010.13760",
    archivePrefix = "arXiv",
    primaryClass = "hep-ph",
    reportNumber = "DESY 20-183, DESY-20-183",
    doi = "10.1103/PhysRevD.104.095027",
    journal = "Phys. Rev. D",
    volume = "104",
    number = "9",
    pages = "095027",
    year = "2021"
}

@article{Harigaya:2023bmp,
    author = "Harigaya, Keisuke and Wang, Isaac R.",
    title = "{ALP-assisted strong first-order electroweak phase transition and baryogenesis}",
    eprint = "2309.00587",
    archivePrefix = "arXiv",
    primaryClass = "hep-ph",
    doi = "10.1007/JHEP04(2024)108",
    journal = "JHEP",
    volume = "04",
    pages = "108",
    year = "2024"
}

@article{DiLuzio:2020wdo,
    author = "Di Luzio, Luca and Giannotti, Maurizio and Nardi, Enrico and Visinelli, Luca",
    title = "{The landscape of QCD axion models}",
    eprint = "2003.01100",
    archivePrefix = "arXiv",
    primaryClass = "hep-ph",
    reportNumber = "DESY 20-036, DESY-20-036",
    doi = "10.1016/j.physrep.2020.06.002",
    journal = "Phys. Rept.",
    volume = "870",
    pages = "1--117",
    year = "2020"
}

@article{Jaeckel:2010ni,
    author = "Jaeckel, Joerg and Ringwald, Andreas",
    title = "{The Low-Energy Frontier of Particle Physics}",
    eprint = "1002.0329",
    archivePrefix = "arXiv",
    primaryClass = "hep-ph",
    reportNumber = "CPT-10-18, DESY-10-016, IPPP-10-09",
    doi = "10.1146/annurev.nucl.012809.104433",
    journal = "Ann. Rev. Nucl. Part. Sci.",
    volume = "60",
    pages = "405--437",
    year = "2010"
}

@article{Graham:2015ouw,
    author = "Graham, Peter W. and Irastorza, Igor G. and Lamoreaux, Steven K. and Lindner, Axel and van Bibber, Karl A.",
    title = "{Experimental Searches for the Axion and Axion-Like Particles}",
    eprint = "1602.00039",
    archivePrefix = "arXiv",
    primaryClass = "hep-ex",
    doi = "10.1146/annurev-nucl-102014-022120",
    journal = "Ann. Rev. Nucl. Part. Sci.",
    volume = "65",
    pages = "485--514",
    year = "2015"
}

@article{Irastorza:2018dyq,
    author = "Irastorza, Igor G. and Redondo, Javier",
    title = "{New experimental approaches in the search for axion-like particles}",
    eprint = "1801.08127",
    archivePrefix = "arXiv",
    primaryClass = "hep-ph",
    doi = "10.1016/j.ppnp.2018.05.003",
    journal = "Prog. Part. Nucl. Phys.",
    volume = "102",
    pages = "89--159",
    year = "2018"
}

@article{Bradley:2003kg,
    author = "Bradley, R. and Clarke, J. and Kinion, D. and Rosenberg, L. J. and van Bibber, K. and Matsuki, S. and Muck, M. and Sikivie, P.",
    title = "{Microwave cavity searches for dark-matter axions}",
    doi = "10.1103/RevModPhys.75.777",
    journal = "Rev. Mod. Phys.",
    volume = "75",
    pages = "777--817",
    year = "2003"
}

@article{ADMX:2019uok,
    author = "Braine, T. and others",
    collaboration = "ADMX",
    title = "{Extended Search for the Invisible Axion with the Axion Dark Matter Experiment}",
    eprint = "1910.08638",
    archivePrefix = "arXiv",
    primaryClass = "hep-ex",
    reportNumber = "FERMILAB-PUB-19-569-AD-AE-PPD",
    doi = "10.1103/PhysRevLett.124.101303",
    journal = "Phys. Rev. Lett.",
    volume = "124",
    number = "10",
    pages = "101303",
    year = "2020"
}

@article{MADMAX:2019pub,
    author = "Brun, P. and others",
    collaboration = "MADMAX",
    title = "{A new experimental approach to probe QCD axion dark matter in the mass range above 40 $\mu$eV}",
    eprint = "1901.07401",
    archivePrefix = "arXiv",
    primaryClass = "physics.ins-det",
    reportNumber = "DESY-19-011, DESY 19-011",
    doi = "10.1140/epjc/s10052-019-6683-x",
    journal = "Eur. Phys. J. C",
    volume = "79",
    number = "3",
    pages = "186",
    year = "2019"
}

@article{IAXO:2019mpb,
    author = "Armengaud, E. and others",
    collaboration = "IAXO",
    title = "{Physics potential of the International Axion Observatory (IAXO)}",
    eprint = "1904.09155",
    archivePrefix = "arXiv",
    primaryClass = "hep-ph",
    doi = "10.1088/1475-7516/2019/06/047",
    journal = "JCAP",
    volume = "06",
    pages = "047",
    year = "2019"
}

@article{IAXO:2020wwp,
    author = "Abeln, A. and others",
    collaboration = "IAXO",
    title = "{Conceptual design of BabyIAXO, the intermediate stage towards the International Axion Observatory}",
    eprint = "2010.12076",
    archivePrefix = "arXiv",
    primaryClass = "physics.ins-det",
    doi = "10.1007/JHEP05(2021)137",
    journal = "JHEP",
    volume = "05",
    pages = "137",
    year = "2021"
}

@article{CAST:2024eil,
    author = {Altenm{\"u}ller, K. and others},
    collaboration = "CAST",
    title = "{New Upper Limit on the Axion-Photon Coupling with an Extended CAST Run with a Xe-Based Micromegas Detector}",
    eprint = "2406.16840",
    archivePrefix = "arXiv",
    primaryClass = "hep-ex",
    doi = "10.1103/PhysRevLett.133.221005",
    journal = "Phys. Rev. Lett.",
    volume = "133",
    number = "22",
    pages = "221005",
    year = "2024"
}

@article{Redondo:2010dp,
    author = "Redondo, Javier and Ringwald, Andreas",
    title = "{Light shining through walls}",
    eprint = "1011.3741",
    archivePrefix = "arXiv",
    primaryClass = "hep-ph",
    reportNumber = "DESY-10-175, MPP-2010-149",
    doi = "10.1080/00107514.2011.563516",
    journal = "Contemp. Phys.",
    volume = "52",
    pages = "211--236",
    year = "2011"
}

@article{ALPSII:2025eri,
    author = "Brotherton, Daniel C. and others",
    collaboration = "ALPS II",
    title = "{Any Light Particle Searches with ALPS II: first science results}",
    eprint = "2512.14110",
    archivePrefix = "arXiv",
    primaryClass = "hep-ex",
    month = "12",
    year = "2025"
}

@article{Mimasu:2014nea,
    author = "Mimasu, Ken and Sanz, Ver{\'o}nica",
    title = "{ALPs at Colliders}",
    eprint = "1409.4792",
    archivePrefix = "arXiv",
    primaryClass = "hep-ph",
    doi = "10.1007/JHEP06(2015)173",
    journal = "JHEP",
    volume = "06",
    pages = "173",
    year = "2015"
}

@article{Jaeckel:2015jla,
    author = "Jaeckel, Joerg and Spannowsky, Michael",
    title = "{Probing MeV to 90 GeV axion-like particles with LEP and LHC}",
    eprint = "1509.00476",
    archivePrefix = "arXiv",
    primaryClass = "hep-ph",
    doi = "10.1016/j.physletb.2015.12.037",
    journal = "Phys. Lett. B",
    volume = "753",
    pages = "482--487",
    year = "2016"
}

@article{Bauer:2017ris,
    author = "Bauer, Martin and Neubert, Matthias and Thamm, Andrea",
    title = "{Collider Probes of Axion-Like Particles}",
    eprint = "1708.00443",
    archivePrefix = "arXiv",
    primaryClass = "hep-ph",
    reportNumber = "MITP-17-047",
    doi = "10.1007/JHEP12(2017)044",
    journal = "JHEP",
    volume = "12",
    pages = "044",
    year = "2017"
}

@article{Bauer:2018uxu,
    author = "Bauer, Martin and Heiles, Mathias and Neubert, Matthias and Thamm, Andrea",
    title = "{Axion-Like Particles at Future Colliders}",
    eprint = "1808.10323",
    archivePrefix = "arXiv",
    primaryClass = "hep-ph",
    reportNumber = "CERN-TH-2018-199, MITP/18-075",
    doi = "10.1140/epjc/s10052-019-6587-9",
    journal = "Eur. Phys. J. C",
    volume = "79",
    number = "1",
    pages = "74",
    year = "2019"
}

@article{Dobrich:2019dxc,
    author = {D{\"o}brich, Babette and Jaeckel, Joerg and Spadaro, Tommaso},
    title = "{Light in the beam dump - ALP production from decay photons in proton beam-dumps}",
    eprint = "1904.02091",
    archivePrefix = "arXiv",
    primaryClass = "hep-ph",
    doi = "10.1007/JHEP05(2019)213",
    journal = "JHEP",
    volume = "05",
    pages = "213",
    year = "2019",
    note = "[Erratum: JHEP 10, 046 (2020)]"
}

@article{NA64:2020qwq,
    author = "Banerjee, D. and others",
    collaboration = "NA64",
    title = "{Search for Axionlike and Scalar Particles with the NA64 Experiment}",
    eprint = "2005.02710",
    archivePrefix = "arXiv",
    primaryClass = "hep-ex",
    reportNumber = "CERN-EP-2020-068",
    doi = "10.1103/PhysRevLett.125.081801",
    journal = "Phys. Rev. Lett.",
    volume = "125",
    number = "8",
    pages = "081801",
    year = "2020"
}

@inproceedings{Tammaro:2025zso,
    author = "Tammaro, Michele and Zupan, Jure",
    title = "{Axion searches at colliders}",
    eprint = "2505.00124",
    archivePrefix = "arXiv",
    primaryClass = "hep-ph",
    month = "4",
    year = "2025"
}

@article{Carenza:2024ehj,
    author = "Carenza, Pierluca and Giannotti, Maurizio and Isern, Jordi and Mirizzi, Alessandro and Straniero, Oscar",
    title = "{Axion astrophysics}",
    eprint = "2411.02492",
    archivePrefix = "arXiv",
    primaryClass = "hep-ph",
    reportNumber = "BARI-TH/66-24",
    doi = "10.1016/j.physrep.2025.02.002",
    journal = "Phys. Rept.",
    volume = "1117",
    pages = "1--102",
    year = "2025"
}

@article{Caputo:2024oqc,
    author = "Caputo, Andrea and Raffelt, Georg",
    title = "{Astrophysical Axion Bounds: The 2024 Edition}",
    eprint = "2401.13728",
    archivePrefix = "arXiv",
    primaryClass = "hep-ph",
    reportNumber = "MPP-2024-13, CERN-TH-2024-013",
    doi = "10.22323/1.454.0041",
    journal = "PoS",
    volume = "COSMICWISPers",
    pages = "041",
    year = "2024"
}

@article{Slatyer:2016qyl,
    author = "Slatyer, Tracy R. and Wu, Chih-Liang",
    title = "{General Constraints on Dark Matter Decay from the Cosmic Microwave Background}",
    eprint = "1610.06933",
    archivePrefix = "arXiv",
    primaryClass = "astro-ph.CO",
    reportNumber = "MIT-CTP-4842",
    doi = "10.1103/PhysRevD.95.023010",
    journal = "Phys. Rev. D",
    volume = "95",
    number = "2",
    pages = "023010",
    year = "2017"
}

@article{Bolliet:2020ofj,
    author = "Bolliet, Boris and Chluba, Jens and Battye, Richard",
    title = "{Spectral distortion constraints on photon injection from low-mass decaying particles}",
    eprint = "2012.07292",
    archivePrefix = "arXiv",
    primaryClass = "astro-ph.CO",
    doi = "10.1093/mnras/stab1997",
    journal = "Mon. Not. Roy. Astron. Soc.",
    volume = "507",
    number = "3",
    pages = "3148--3178",
    year = "2021"
}

@article{Horiuchi:2013noa,
    author = "Horiuchi, Shunsaku and Humphrey, Philip J. and Onorbe, Jose and Abazajian, Kevork N. and Kaplinghat, Manoj and Garrison-Kimmel, Shea",
    title = "{Sterile neutrino dark matter bounds from galaxies of the Local Group}",
    eprint = "1311.0282",
    archivePrefix = "arXiv",
    primaryClass = "astro-ph.CO",
    doi = "10.1103/PhysRevD.89.025017",
    journal = "Phys. Rev. D",
    volume = "89",
    number = "2",
    pages = "025017",
    year = "2014"
}

@article{Foster:2022ajl,
    author = "Foster, Joshua W. and Kumar, Soubhik and Safdi, Benjamin R. and Soreq, Yotam",
    title = "{Dark Grand Unification in the axiverse: decaying axion dark matter and spontaneous baryogenesis}",
    eprint = "2208.10504",
    archivePrefix = "arXiv",
    primaryClass = "hep-ph",
    reportNumber = "MIT-CTP/5458",
    doi = "10.1007/JHEP12(2022)119",
    journal = "JHEP",
    volume = "12",
    pages = "119",
    year = "2022"
}

@article{Roach:2022lgo,
    author = "Roach, Brandon M. and Rossland, Steven and Ng, Kenny C. Y. and Perez, Kerstin and Beacom, John F. and Grefenstette, Brian W. and Horiuchi, Shunsaku and Krivonos, Roman and Wik, Daniel R.",
    title = "{Long-exposure NuSTAR constraints on decaying dark matter in the Galactic halo}",
    eprint = "2207.04572",
    archivePrefix = "arXiv",
    primaryClass = "astro-ph.HE",
    doi = "10.1103/PhysRevD.107.023009",
    journal = "Phys. Rev. D",
    volume = "107",
    number = "2",
    pages = "023009",
    year = "2023"
}

@article{Laha:2020ivk,
    author = "Laha, Ranjan and Mu{\~n}oz, Julian B. and Slatyer, Tracy R.",
    title = "{INTEGRAL constraints on primordial black holes and particle dark matter}",
    eprint = "2004.00627",
    archivePrefix = "arXiv",
    primaryClass = "astro-ph.CO",
    doi = "10.1103/PhysRevD.101.123514",
    journal = "Phys. Rev. D",
    volume = "101",
    number = "12",
    pages = "123514",
    year = "2020"
}

@article{Panci:2022wlc,
    author = "Panci, Paolo and Redigolo, Diego and Schwetz, Thomas and Ziegler, Robert",
    title = "{Axion dark matter from lepton flavor-violating decays}",
    eprint = "2209.03371",
    archivePrefix = "arXiv",
    primaryClass = "hep-ph",
    doi = "10.1016/j.physletb.2023.137919",
    journal = "Phys. Lett. B",
    volume = "841",
    pages = "137919",
    year = "2023"
}

@article{Leutwyler:1989tn,
    author = "Leutwyler, H. and Shifman, Mikhail A.",
    title = "{GOLDSTONE BOSONS GENERATE PECULIAR CONFORMAL ANOMALIES}",
    reportNumber = "BUTP-89/02-BERN",
    doi = "10.1016/0370-2693(89)91730-9",
    journal = "Phys. Lett. B",
    volume = "221",
    pages = "384--388",
    year = "1989"
}

@article{GrillidiCortona:2015jxo,
    author = "Grilli di Cortona, Giovanni and Hardy, Edward and Pardo Vega, Javier and Villadoro, Giovanni",
    title = "{The QCD axion, precisely}",
    eprint = "1511.02867",
    archivePrefix = "arXiv",
    primaryClass = "hep-ph",
    doi = "10.1007/JHEP01(2016)034",
    journal = "JHEP",
    volume = "01",
    pages = "034",
    year = "2016"
}

@article{Fradette:2018hhl,
    author = "Fradette, Anthony and Pospelov, Maxim and Pradler, Josef and Ritz, Adam",
    title = "{Cosmological beam dump: constraints on dark scalars mixed with the Higgs boson}",
    eprint = "1812.07585",
    archivePrefix = "arXiv",
    primaryClass = "hep-ph",
    doi = "10.1103/PhysRevD.99.075004",
    journal = "Phys. Rev. D",
    volume = "99",
    number = "7",
    pages = "075004",
    year = "2019"
}

@article{Lu:2020rhp,
    author = "Lu, Zhen-Yan and Du, Meng-Lin and Guo, Feng-Kun and Mei{\ss}ner, Ulf-G. and Vonk, Thomas",
    title = "{QCD $\theta$-vacuum energy and axion properties}",
    eprint = "2003.01625",
    archivePrefix = "arXiv",
    primaryClass = "hep-ph",
    doi = "10.1007/JHEP05(2020)001",
    journal = "JHEP",
    volume = "05",
    pages = "001",
    year = "2020"
}

@article{Bauer:2020jbp,
    author = "Bauer, Martin and Neubert, Matthias and Renner, Sophie and Schnubel, Marvin and Thamm, Andrea",
    title = "{The Low-Energy Effective Theory of Axions and ALPs}",
    eprint = "2012.12272",
    archivePrefix = "arXiv",
    primaryClass = "hep-ph",
    reportNumber = "IPPP/20/69, MITP/20-070 SISSA 30/2020/FISI, ZH-TH-47/20",
    doi = "10.1007/JHEP04(2021)063",
    journal = "JHEP",
    volume = "04",
    pages = "063",
    year = "2021"
}

@article{Ertas:2020xcc,
    author = "Ertas, Fatih and Kahlhoefer, Felix",
    title = "{On the interplay between astrophysical and laboratory probes of MeV-scale axion-like particles}",
    eprint = "2004.01193",
    archivePrefix = "arXiv",
    primaryClass = "hep-ph",
    reportNumber = "TTK-20-08",
    doi = "10.1007/JHEP07(2020)050",
    journal = "JHEP",
    volume = "07",
    pages = "050",
    year = "2020"
}

@article{DallaValleGarcia:2023xhh,
    author = "Dalla Valle Garcia, Giovani and Kahlhoefer, Felix and Ovchynnikov, Maksym and Zaporozhchenko, Andrii",
    title = "{Phenomenology of axionlike particles with universal fermion couplings revisited}",
    eprint = "2310.03524",
    archivePrefix = "arXiv",
    primaryClass = "hep-ph",
    reportNumber = "TTP23-042, P3H-23-070",
    doi = "10.1103/PhysRevD.109.055042",
    journal = "Phys. Rev. D",
    volume = "109",
    number = "5",
    pages = "055042",
    year = "2024"
}

@article{Aghaie:2024jkj,
    author = "Aghaie, Mohammad and Armando, Giovanni and Conaci, Angela and Dondarini, Alessandro and Matak, Peter and Panci, Paolo and Sinska, Zuzana and Ziegler, Robert",
    title = "{Axion dark matter from heavy quarks}",
    eprint = "2404.12199",
    archivePrefix = "arXiv",
    primaryClass = "hep-ph",
    doi = "10.1016/j.physletb.2024.138923",
    journal = "Phys. Lett. B",
    volume = "856",
    pages = "138923",
    year = "2024"
}

@article{Bai:2024lpq,
    author = "Bai, Yang and Chen, Ting-Kuo and Liu, Jia and Ma, Xiaolin",
    title = "{Wess-Zumino-Witten Interactions of Axions}",
    eprint = "2406.11948",
    archivePrefix = "arXiv",
    primaryClass = "hep-ph",
    doi = "10.1103/PhysRevLett.134.081803",
    journal = "Phys. Rev. Lett.",
    volume = "134",
    number = "8",
    pages = "081803",
    year = "2025"
}

@article{Flambaum:2024zyt,
    author = "Flambaum, V. V. and Samsonov, I. B.",
    title = "{Limits on scalar dark matter interactions with particles other than the photon via loop corrections to the scalar-photon coupling}",
    eprint = "2403.02685",
    archivePrefix = "arXiv",
    primaryClass = "hep-ph",
    doi = "10.1103/PhysRevD.110.075044",
    journal = "Phys. Rev. D",
    volume = "110",
    number = "7",
    pages = "075044",
    year = "2024"
}

@article{Bai:2025fvl,
    author = "Bai, Yang and Chen, Ting-Kuo and Liu, Jia and Ma, Xiaolin",
    title = "{Wess-Zumino-Witten Interactions of Axions: Three-Flavor}",
    eprint = "2505.24822",
    archivePrefix = "arXiv",
    primaryClass = "hep-ph",
    month = "5",
    year = "2025"
}

@article{Delaunay:2025lhl,
    author = "Delaunay, C{\'e}dric and Kitahara, Teppei and Soreq, Yotam and Zupan, Jure",
    title = "{Light scalar beyond the Higgs mixing limit}",
    eprint = "2501.16477",
    archivePrefix = "arXiv",
    primaryClass = "hep-ph",
    doi = "10.1007/JHEP10(2025)222",
    journal = "JHEP",
    volume = "10",
    pages = "222",
    year = "2025"
}

@article{Chala:2020wvs,
    author = "Chala, Mikael and Guedes, Guilherme and Ramos, Maria and Santiago, Jose",
    title = "{Running in the ALPs}",
    eprint = "2012.09017",
    archivePrefix = "arXiv",
    primaryClass = "hep-ph",
    doi = "10.1140/epjc/s10052-021-08968-2",
    journal = "Eur. Phys. J. C",
    volume = "81",
    number = "2",
    pages = "181",
    year = "2021"
}

@article{Bonilla:2021ufe,
    author = "Bonilla, J. and Brivio, I. and Gavela, M. B. and Sanz, V.",
    title = "{One-loop corrections to ALP couplings}",
    eprint = "2107.11392",
    archivePrefix = "arXiv",
    primaryClass = "hep-ph",
    reportNumber = "IFT-UAM/CSIC-21-82",
    doi = "10.1007/JHEP11(2021)168",
    journal = "JHEP",
    volume = "11",
    pages = "168",
    year = "2021"
}

@article{DasBakshi:2023lca,
    author = "Das Bakshi, Supratim and Machado-Rodr{\'\i}guez, Jonathan and Ramos, Maria",
    title = "{Running beyond ALPs: shift-breaking and CP-violating effects}",
    eprint = "2306.08036",
    archivePrefix = "arXiv",
    primaryClass = "hep-ph",
    reportNumber = "IFT-UAM/CSIC-23-64",
    doi = "10.1007/JHEP11(2023)133",
    journal = "JHEP",
    volume = "11",
    pages = "133",
    year = "2023"
}

@article{DiLuzio:2023lmd,
    author = "Di Luzio, Luca and Gisbert, Hector and Levati, Gabriele and Paradisi, Paride and S{\o}rensen, Philip",
    title = "{CP-Violating Axions: A Theory Review}",
    eprint = "2312.17310",
    archivePrefix = "arXiv",
    primaryClass = "hep-ph",
    month = "12",
    year = "2023"
}

@article{Bresciani:2024shu,
    author = "Bresciani, Luigi C. and Brunello, Giacomo and Levati, Gabriele and Mastrolia, Pierpaolo and Paradisi, Paride",
    title = "{Renormalization of effective field theories via on-shell methods: the case of axion-like particles}",
    eprint = "2412.04160",
    archivePrefix = "arXiv",
    primaryClass = "hep-ph",
    doi = "10.1007/JHEP10(2025)190",
    journal = "JHEP",
    volume = "10",
    pages = "190",
    year = "2025"
}

@article{Misiak:2025xzq,
    author = "Misiak, Miko{\l}aj and Na{\l}{\k{e}}cz, Ignacy",
    title = "{One-loop renormalization group equations in generic effective field theories. Part I. Bosonic operators}",
    eprint = "2501.17134",
    archivePrefix = "arXiv",
    primaryClass = "hep-ph",
    doi = "10.1007/JHEP06(2025)210",
    journal = "JHEP",
    volume = "06",
    pages = "210",
    year = "2025"
}

@article{Fonseca:2025zjb,
    author = "Fonseca, Renato M. and Olgoso, Pablo and Santiago, Jos{\'e}",
    title = "{Renormalization of general Effective Field Theories: formalism and renormalization of bosonic operators}",
    eprint = "2501.13185",
    archivePrefix = "arXiv",
    primaryClass = "hep-ph",
    doi = "10.1007/JHEP07(2025)135",
    journal = "JHEP",
    volume = "07",
    pages = "135",
    year = "2025"
}

@article{Aebischer:2025zxg,
    author = "Aebischer, Jason and Bresciani, Luigi C. and Selimovic, Nudzeim",
    title = "{Anomalous dimension of a general effective gauge theory. Part I. Bosonic sector}",
    eprint = "2502.14030",
    archivePrefix = "arXiv",
    primaryClass = "hep-ph",
    reportNumber = "CERN-TH-2025-032",
    doi = "10.1007/JHEP08(2025)209",
    journal = "JHEP",
    volume = "08",
    pages = "209",
    year = "2025"
}

@phdthesis{Levati:2024pvl,
    author = "Levati, Gabriele",
    title = "{Effective Field Theory tools for investigating physics Beyond the Standard Model}",
    school = "Universit{\`a} degli studi di Padova, Italy, U. Padua (main)",
    year = "2024"
}

@article{Georgi:1986df,
    author = "Georgi, Howard and Kaplan, David B. and Randall, Lisa",
    title = "{Manifesting the Invisible Axion at Low-energies}",
    reportNumber = "HUTP-86/A004",
    doi = "10.1016/0370-2693(86)90688-X",
    journal = "Phys. Lett. B",
    volume = "169",
    pages = "73--78",
    year = "1986"
}

@article{DiLuzio:2023cuk,
    author = "Di Luzio, Luca and Levati, Gabriele and Paradisi, Paride",
    title = "{The chiral Lagrangian of CP-violating axion-like particles}",
    eprint = "2311.12158",
    archivePrefix = "arXiv",
    primaryClass = "hep-ph",
    doi = "10.1007/JHEP02(2024)020",
    journal = "JHEP",
    volume = "02",
    pages = "020",
    year = "2024"
}

@article{Wess:1971yu,
    author = "Wess, J. and Zumino, B.",
    title = "{Consequences of anomalous Ward identities}",
    doi = "10.1016/0370-2693(71)90582-X",
    journal = "Phys. Lett. B",
    volume = "37",
    pages = "95--97",
    year = "1971"
}

@article{Witten:1983tw,
    author = "Witten, Edward",
    title = "{Global Aspects of Current Algebra}",
    reportNumber = "PRINT-83-0262 (PRINCETON)",
    doi = "10.1016/0550-3213(83)90063-9",
    journal = "Nucl. Phys. B",
    volume = "223",
    pages = "422--432",
    year = "1983"
}

@article{ParticleDataGroup:2024cfk,
    author = "Navas, S. and others",
    collaboration = "Particle Data Group",
    title = "{Review of particle physics}",
    doi = "10.1103/PhysRevD.110.030001",
    journal = "Phys. Rev. D",
    volume = "110",
    number = "3",
    pages = "030001",
    year = "2024"
}

@article{Vecchi:2010gj,
    author = "Vecchi, Luca",
    title = "{Phenomenology of a light scalar: the dilaton}",
    eprint = "1002.1721",
    archivePrefix = "arXiv",
    primaryClass = "hep-ph",
    doi = "10.1103/PhysRevD.82.076009",
    journal = "Phys. Rev. D",
    volume = "82",
    pages = "076009",
    year = "2010"
}

@article{Chacko:2012sy,
    author = "Chacko, Zackaria and Mishra, Rashmish K.",
    title = "{Effective Theory of a Light Dilaton}",
    eprint = "1209.3022",
    archivePrefix = "arXiv",
    primaryClass = "hep-ph",
    doi = "10.1103/PhysRevD.87.115006",
    journal = "Phys. Rev. D",
    volume = "87",
    number = "11",
    pages = "115006",
    year = "2013"
}

@article{Ahmed:2019csf,
    author = "Ahmed, Aqeel and Mariotti, Alberto and Najjari, Saereh",
    title = "{A light dilaton at the LHC}",
    eprint = "1912.06645",
    archivePrefix = "arXiv",
    primaryClass = "hep-ph",
    doi = "10.1007/JHEP05(2020)093",
    journal = "JHEP",
    volume = "05",
    pages = "093",
    year = "2020"
}

@article{Appelquist:2022mjb,
    author = "Appelquist, Thomas and Ingoldby, James and Piai, Maurizio",
    title = "{Dilaton Effective Field Theory}",
    eprint = "2209.14867",
    archivePrefix = "arXiv",
    primaryClass = "hep-ph",
    doi = "10.3390/universe9010010",
    journal = "Universe",
    volume = "9",
    number = "1",
    pages = "10",
    year = "2023"
}

@article{Bellazzini:2013fga,
    author = "Bellazzini, Brando and Csaki, Csaba and Hubisz, Jay and Serra, Javi and Terning, John",
    title = "{A Naturally Light Dilaton and a Small Cosmological Constant}",
    eprint = "1305.3919",
    archivePrefix = "arXiv",
    primaryClass = "hep-th",
    doi = "10.1140/epjc/s10052-014-2790-x",
    journal = "Eur. Phys. J. C",
    volume = "74",
    pages = "2790",
    year = "2014"
}

@article{Coradeschi:2013gda,
    author = "Coradeschi, Francesco and Lodone, Paolo and Pappadopulo, Duccio and Rattazzi, Riccardo and Vitale, Lorenzo",
    title = "{A naturally light dilaton}",
    eprint = "1306.4601",
    archivePrefix = "arXiv",
    primaryClass = "hep-th",
    doi = "10.1007/JHEP11(2013)057",
    journal = "JHEP",
    volume = "11",
    pages = "057",
    year = "2013"
}

@article{Abu-Ajamieh:2017khi,
    author = "Abu-Ajamieh, Fayez and Lee, Jun Seok and Terning, John",
    title = "{The Light Radion Window}",
    eprint = "1711.02697",
    archivePrefix = "arXiv",
    primaryClass = "hep-ph",
    doi = "10.1007/JHEP10(2018)050",
    journal = "JHEP",
    volume = "10",
    pages = "050",
    year = "2018"
}

@article{Megias:2015qqh,
    author = "Megias, Eugenio and Pujolas, Oriol and Quiros, Mariano",
    editor = "Bravina, Larisa and Foka, Yiota and Kabana, Sonja",
    title = "{On light dilaton extensions of the Standard Model}",
    eprint = "1512.06702",
    archivePrefix = "arXiv",
    primaryClass = "hep-ph",
    reportNumber = "MPP-2015-272, UAB-FT-769",
    doi = "10.1051/epjconf/201612605010",
    journal = "EPJ Web Conf.",
    volume = "126",
    pages = "05010",
    year = "2016"
}

@article{Agrawal:2016ubh,
    author = "Agrawal, Prateek and Sundrum, Raman",
    title = "{Small Vacuum Energy from Small Equivalence Violation in Scalar Gravity}",
    eprint = "1611.07021",
    archivePrefix = "arXiv",
    primaryClass = "hep-th",
    reportNumber = "UMD-PP-017-012",
    doi = "10.1007/JHEP05(2017)144",
    journal = "JHEP",
    volume = "05",
    pages = "144",
    year = "2017"
}

@article{Dolan:2014ska,
    author = "Dolan, Matthew J. and Kahlhoefer, Felix and McCabe, Christopher and Schmidt-Hoberg, Kai",
    title = "{A taste of dark matter: Flavour constraints on pseudoscalar mediators}",
    eprint = "1412.5174",
    archivePrefix = "arXiv",
    primaryClass = "hep-ph",
    reportNumber = "DESY-14-238, SLAC-PUB-16179",
    doi = "10.1007/JHEP03(2015)171",
    journal = "JHEP",
    volume = "03",
    pages = "171",
    year = "2015",
    note = "[Erratum: JHEP 07, 103 (2015)]"
}

@article{Nelson:1983zb,
    author = "Nelson, Ann E.",
    title = "{Naturally Weak CP Violation}",
    reportNumber = "HUTP-83/A080",
    doi = "10.1016/0370-2693(84)92025-2",
    journal = "Phys. Lett. B",
    volume = "136",
    pages = "387--391",
    year = "1984"
}

@article{Barr:1984qx,
    author = "Barr, Stephen M.",
    title = "{Solving the Strong CP Problem Without the Peccei-Quinn Symmetry}",
    reportNumber = "DOE-ER-40048-08 P4",
    doi = "10.1103/PhysRevLett.53.329",
    journal = "Phys. Rev. Lett.",
    volume = "53",
    pages = "329",
    year = "1984"
}

@article{Feruglio:2024dnc,
    author = "Feruglio, Ferruccio and Ziegler, Robert",
    title = "{CPon Dark Matter}",
    eprint = "2411.08101",
    archivePrefix = "arXiv",
    primaryClass = "hep-ph",
    reportNumber = "TTP24-043, P3H-24-078",
    doi = "10.1007/JHEP03(2025)102",
    journal = "JHEP",
    volume = "03",
    pages = "102",
    year = "2025"
}

@article{Feruglio:2023uof,
    author = "Feruglio, Ferruccio and Strumia, Alessandro and Titov, Arsenii",
    title = "{Modular invariance and the QCD angle}",
    eprint = "2305.08908",
    archivePrefix = "arXiv",
    primaryClass = "hep-ph",
    doi = "10.1007/JHEP07(2023)027",
    journal = "JHEP",
    volume = "07",
    pages = "027",
    year = "2023"
}

@article{Feruglio:2024ytl,
    author = "Feruglio, Ferruccio and Parriciatu, Matteo and Strumia, Alessandro and Titov, Arsenii",
    title = "{Solving the strong CP problem without axions}",
    eprint = "2406.01689",
    archivePrefix = "arXiv",
    primaryClass = "hep-ph",
    doi = "10.1007/JHEP08(2024)214",
    journal = "JHEP",
    volume = "08",
    pages = "214",
    year = "2024"
}

@article{Georgi:1993jn,
    author = "Georgi, Howard",
    title = "{A bound on m(eta) / m(eta-prime) for large n(c)}",
    eprint = "hep-ph/9310337",
    archivePrefix = "arXiv",
    reportNumber = "HUTP-93-A029",
    doi = "10.1103/PhysRevD.49.1666",
    journal = "Phys. Rev. D",
    volume = "49",
    pages = "1666--1667",
    year = "1994"
}

@article{Gerard:2004gx,
    author = "Gerard, J. -M. and Kou, E.",
    title = "{eta-eta-prime masses and mixing: A Large N(c) reappraisal}",
    eprint = "hep-ph/0411292",
    archivePrefix = "arXiv",
    reportNumber = "UCL-IPT-04-21",
    doi = "10.1016/j.physletb.2005.04.057",
    journal = "Phys. Lett. B",
    volume = "616",
    pages = "85--92",
    year = "2005"
}

@article{Leutwyler:1997yr,
    author = "Leutwyler, H.",
    editor = "Narison, Stephan",
    title = "{On the 1/N expansion in chiral perturbation theory}",
    eprint = "hep-ph/9709408",
    archivePrefix = "arXiv",
    doi = "10.1016/S0920-5632(97)01065-7",
    journal = "Nucl. Phys. B Proc. Suppl.",
    volume = "64",
    pages = "223--231",
    year = "1998"
}

@article{Beisert:2001qb,
    author = "Beisert, Niklas and Borasoy, Bugra",
    title = "{eta eta-prime mixing in U(3) chiral perturbation theory}",
    eprint = "hep-ph/0107175",
    archivePrefix = "arXiv",
    doi = "10.1007/s100500170072",
    journal = "Eur. Phys. J. A",
    volume = "11",
    pages = "329--339",
    year = "2001"
}

@article{Alves:2017avw,
    author = "Alves, Daniele S. M. and Weiner, Neal",
    title = "{A viable QCD axion in the MeV mass range}",
    eprint = "1710.03764",
    archivePrefix = "arXiv",
    primaryClass = "hep-ph",
    reportNumber = "LA-UR-17-29295",
    doi = "10.1007/JHEP07(2018)092",
    journal = "JHEP",
    volume = "07",
    pages = "092",
    year = "2018"
}

@article{Aloni:2018vki,
    author = "Aloni, Daniel and Soreq, Yotam and Williams, Mike",
    title = "{Coupling QCD-Scale Axionlike Particles to Gluons}",
    eprint = "1811.03474",
    archivePrefix = "arXiv",
    primaryClass = "hep-ph",
    reportNumber = "CERN-TH-2018-237, MIT-CTP/5080, MIT-CTP-5080",
    doi = "10.1103/PhysRevLett.123.031803",
    journal = "Phys. Rev. Lett.",
    volume = "123",
    number = "3",
    pages = "031803",
    year = "2019"
}

@article{Cheng:2021kjg,
    author = "Cheng, Hsin-Chia and Li, Lingfeng and Salvioni, Ennio",
    title = "{A theory of dark pions}",
    eprint = "2110.10691",
    archivePrefix = "arXiv",
    primaryClass = "hep-ph",
    reportNumber = "CERN-TH-2021-150",
    doi = "10.1007/JHEP01(2022)122",
    journal = "JHEP",
    volume = "01",
    pages = "122",
    year = "2022"
}

@article{Ovchynnikov:2025gpx,
    author = "Ovchynnikov, Maksym and Zaporozhchenko, Andrii",
    title = "{Advancing the phenomenology of GeV-scale axionlike particles}",
    eprint = "2501.04525",
    archivePrefix = "arXiv",
    primaryClass = "hep-ph",
    reportNumber = "CERN-TH-2025-006",
    doi = "10.1103/568p-d1ls",
    journal = "Phys. Rev. D",
    volume = "112",
    number = "1",
    pages = "015001",
    year = "2025"
}

@article{Arina:2021nqi,
    author = "Arina, Chiara and Hajer, Jan and Klose, Philipp",
    title = "{Portal Effective Theories. A framework for the model independent description of light hidden sector interactions}",
    eprint = "2105.06477",
    archivePrefix = "arXiv",
    primaryClass = "hep-ph",
    doi = "10.1007/JHEP09(2021)063",
    journal = "JHEP",
    volume = "09",
    pages = "063",
    year = "2021"
}

@article{Leutwyler:1989xj,
    author = "Leutwyler, H. and Shifman, Mikhail A.",
    title = "{Light Higgs Particle in Decays of $K$ and $\eta$ Mesons}",
    reportNumber = "BUTP-89/29-BERN",
    doi = "10.1016/0550-3213(90)90475-S",
    journal = "Nucl. Phys. B",
    volume = "343",
    pages = "369--397",
    year = "1990"
}
\end{small}

\end{document}